\documentclass[12pt,preprint]{aastex}

\def\ltsima{$\; \buildrel < \over \sim \;$}

\def\lsim{\lower.5ex\hbox{\ltsima}}
\def\gtsima{$\; \buildrel > \over \sim \;$}
\def\gsim{\lower.5ex\hbox{\gtsima}}
\begin{document}
\title{Reionization of Hydrogen and Helium by Early Stars and Quasars}

\author{J. Stuart B. Wyithe\altaffilmark{1,3} and Abraham
Loeb\altaffilmark{1,2,4}}

\email{swyithe@cfa.harvard.edu; loeb@sns.ias.edu}

\altaffiltext{1}{Harvard-Smithsonian Center for Astrophysics, 60 Garden
St., Cambridge, MA 02138}

\altaffiltext{2}{Institute for Advanced Study, Princeton, NJ 08540}

\altaffiltext{3}{Hubble Fellow} 

\altaffiltext{4}{Guggenheim Fellow; on leave from the Astronomy Department,
Harvard University}

\begin{abstract}
\noindent 
We compute the reionization histories of hydrogen and helium due to the
ionizing radiation fields produced by stars and quasars.  For the quasars
we use a model based on halo-merger rates that reproduces all known
properties of the quasar luminosity function at high redshifts.  The less
constrained properties of the ionizing radiation produced by stars are
modeled with two free parameters: (i) a transition redshift, $z_{\rm
tran}$, above which the stellar population is dominated by massive,
zero-metallicity stars and below which it is dominated by a Scalo mass
function; (ii) the product of the escape fraction of stellar ionizing
photons from their host galaxies and the star-formation efficiency, $f_{\rm
esc}f_{\star}$.  We constrain the allowed range of these free parameters at
high redshifts based on the lack of the HI Gunn-Peterson trough at $z\la 6$
and the upper limit on the total intergalactic optical depth for electron
scattering, $\tau_{\rm es}<0.18$, from recent cosmic microwave background
(CMB) experiments.  We find that quasars ionize helium by a redshift $z\sim
4$, but cannot reionize hydrogen by themselves before $z\sim 6$.
A major fraction of the allowed combinations of $f_{\rm
esc}f_{\star}$ and $z_{\rm tran}$ lead to an early peak in the ionized
fraction due to metal-free stars at high redshifts.  This sometimes results
in two reionization epochs, namely an early HII or HeIII overlap phase
followed by recombination and a second overlap phase. Even if early overlap
is not achieved, the peak in the visibility function for scattering of the
CMB often coincides with the early ionization phase rather than with the
actual reionization epoch. Consequently, $\tau_{\rm es}$ does not correspond
directly to the reionization redshift. We generically find values of 
$\tau_{\rm es}\ga 7\%$, that should be detectable by the MAP satellite.

\end{abstract}

\keywords{Cosmology: theory -- Early universe -- intergalactic medium --
stars:formation}

\section{Introduction}

Following cosmological recombination at a redshift $z\sim 10^3$, the
baryonic gas filling up the universe became predominantly neutral.  Given
that this gas is known to be mostly ionized today, one arrives at two of
the major questions in current extragalactic astronomy: {\it (i) when were
the cosmic hydrogen and helium re-ionized?} and {\it (ii) which sources
dominated this reionization process?} The answers to both questions are
likely to be different for hydrogen and helium (see review by Barkana \&
Loeb 2001). Recent observations provide preliminary answers to the first
question.  The absorption spectra of SDSS quasars at $z\sim 6$ indicate
that the neutral fraction of hydrogen increases significantly at $z\ga 6$
(Becker et al. 2001; Fan et al. 2002), and the UV spectrum of quasars
implies that helium is fully ionized only at $z\la 3$ (Jacobsen et al 1994;
Tytler 1995; Davidsen et al. 1996; Hogan et al. 1997; Reimers et al. 1997;
Heap et al. 2000; Kriss et al. 2001; Smette et al. 2002).  The latter
observations also indicate, through a cross-correlation between the hydrogen
and helium forest of absorption lines (Kriss et al. 2001; Smette et
al. 2002), that the ionization of helium at $z\sim 3$ had significant
contributions from both quasars and stars. No analogous evidence exists for
hydrogen at $z\ga 6$.

In this paper we make theoretical predictions for the reionization
histories of hydrogen and helium. We calculate the contribution from
quasars using a model that matches all the existing observational data on
the quasar luminosity function at high redshifts (see Wyithe \& Loeb 2002
for details).  We model the less constrained stellar contribution using two
free parameters, which are in turn constrained by existing observational
data. Since the recombination times of both hydrogen and helium are shorter
than the age of the universe in overdense regions of the intergalactic
medium (IGM), it is possible that these species experienced more than one
epoch of reionization.  One of the goals of our detailed study is to
identify the parameter values for which multiple reionization epochs are
possible.

The reionization history has important implications for the temperature and
polarization anisotropies of the cosmic microwave background (CMB).
Anisotropies on scales smaller than the angular size of the horizon at
hydrogen reionization are suppressed by a factor $\sim e^{-\tau_{\rm es}}$
where $\tau_{\rm es}$ is the line-of-sight optical depth for electron
scattering (Hu \& White 1997; Haiman \& Loeb 1997), and secondary
anisotropies are added (Hu 2000 and references therein).  Reionization is
the primary source of polarization anisotropies on large angular scales
(Zaldarriaga \& Seljak 1997; Kamionkowski et al. 1997; Hu 2000). 
Polarization anisotropies might be detected in the near future by MAP on
large angular scales (Kaplinghat 2002) or by ground-based experiments on
small angular scales. So far, the available data on the temperature
anisotropies of the CMB provides an upper limit of $\tau_{\rm es}\la 0.18$
(Wang, Tegmark \& Zaldarriaga 2002; Bond et al. 2002).  We will use this
upper limit to constrain the free parameters of our model.

Throughout the paper we assume density parameters values of
$\Omega_{m}=0.35$ in matter, $\Omega_{b}=0.052$ in baryons,
$\Omega_\Lambda=0.65$ in a cosmological constant, and a Hubble constant of
$H_0=65~{\rm km\,s^{-1}\,Mpc^{-1}}$ (or equivalently $h=0.65$). For
calculations of the Press-Schechter~(1974) mass function (with the
modification of Sheth \& Tormen~1999) we assume a primordial power-spectrum
with a power-law index $n=1$ and the fitting formula to the exact transfer
function of Cold Dark Matter, given by Bardeen et al.~(1986).  Unless
otherwise noted we adopt an rms amplitude of $\sigma_8=0.87$ for mass
density fluctuations in a sphere of radius $8h^{-1}$Mpc.

\section{Reionization in a Clumpy Universe}

The simplest estimate of the epoch of reionization is based on the
following simple considerations.  Given a co-moving density of ionizing
photons $n_\gamma$ in a homogeneous but clumpy medium of comoving density
$n_0$ (where the size of the HII region is much larger than the scale
length of clumpiness), the evolution of the volume filling factor $Q_{\rm
i}$ of ionized regions is (Haiman \& Loeb~1997; Madau et al.~1999; Barkana
\& Loeb~2001)
\begin{equation}
\label{fillfactor}
\frac{dQ_{\rm i}}{dz} = \frac{1}{n_0}\frac{dn_\gamma}{dz}-\alpha_{\rm
B}\frac{C}{a^3}Q_{\rm i}n_{\rm e}\frac{dt}{dz},
\end{equation}
where $\alpha_{\rm B}$ is the case B recombination coefficient, $a=1/(1+z)$ is the
scale factor, $n_{\rm e}$ is the comoving electron density, and
$C\equiv {\langle n_0^2\rangle}/{\langle n_0\rangle^2}$ is the clumping
factor.  This equation describes statistically the transition from a fully
neutral universe to a fully ionized one, and yields reionization redshifts
for hydrogen of around 7-12. Large uncertainties arise in both the source
term and in the value of the clumping factor (because more rapid
recombinations lead to a slower evolution of $Q_{\rm i}$).

A more realistic description of reionization in a clumpy medium is provided
by the model of Miralda-Escude et al.~(2000). In what follows, we draw
primarily from their prescription and refer the reader to the original paper
for a detailed discussion of its motivations and assumptions. The model
assumes that reionization progresses rapidly through islands of lower
density prior to the overlap of individual cosmological ionized
regions. Following overlap, the remaining regions of high density are
gradually ionized. It is therefore hypothesized that at any time, regions
with gas below some critical overdensity $\Delta_{\rm i}\equiv
{\rho_{i}}/{\langle\rho\rangle}$ are ionized while regions of higher
density are not. The assumption of homogeneity in
equation~(\ref{fillfactor}) implies that the volume filling factor equals
the mass filling factor. Therefore, within the model of Miralda-Escude et
al.~(2000) we replace ${dQ_{\rm i}}/{dz}$ by ${dF_{\rm M}(\Delta_{\rm
i})}/{dz}$, where
\begin{equation}
F_{\rm M}(\Delta_{\rm i})=\int_{0}^{\Delta_{\rm i}}d\Delta P_{\rm
V}(\Delta)\Delta
\end{equation}
is the fraction of mass in regions with overdensity below $\Delta_{\rm i}$,
and $P_{\rm V}(\Delta)$ is the volume weighted probability distribution for
$\Delta$. The product $Q_{\rm i}C$ can be rewritten
\begin{equation}
R(\Delta_{\rm i})\equiv Q_{\rm i}C\equiv Q_{\rm i}\frac{\langle
\rho^2\rangle}{\langle \rho\rangle^2} = F(\Delta_{\rm
i})\frac{\int_{0}^{\Delta_{\rm i}}d\Delta P_{\rm
V}(\Delta)\Delta^2}{\int_{0}^{\Delta_{\rm i}}d\Delta P_{\rm V}(\Delta)}
=\int_{0}^{\Delta_{\rm i}}d\Delta P_{\rm V}(\Delta)\Delta^2,
\end{equation}
where $F(\Delta_{\rm i})$ is the fraction of the volume with
$\Delta<\Delta_{\rm i}$. The quantity $\alpha_{\rm B}(1+z)^3n_{\rm
e}\int_{0}^{\Delta_{\rm i}}d\Delta P_{\rm V}(\Delta)\Delta^2$ is therefore
the number of recombinations per atom in the IGM per second. Note that the
term analogous to the clumping factor is calculated from the volume
weighted distribution. The mass fraction $F_{\rm M}(\Delta_{\rm i})$ (or
equivalently $\Delta_{\rm i}$) therefore evolves according to the equation
\begin{equation}
\label{postoverlap}
\frac{dF_{\rm M}(\Delta_{\rm i})}{dz} =
\frac{1}{n_0}\frac{dn_\gamma}{dz}-\alpha_{\rm B}\frac{R(\Delta_{\rm
i})}{a^3}n_{\rm e}\frac{dt}{dz}.
\end{equation}
This equation assumes that all ionizing photons are absorbed shortly after
being emitted, so that there is no background ionizing field, and no loss
of ionizing photons due to redshift. We therefore implicitly assume that 
the mean free path of ionizing photons is much smaller than the Hubble length.
This should be valid at redshifts not too much smaller than the overlap redshift.

The integration of equation~(\ref{postoverlap}) requires knowledge of 
$P_{\rm V}(\Delta)$. 
Miralda-Escude et al.~(2000) found that a good fit to the volume weighted
probability distribution for the density as seen in N-body simulations
has the functional form
\begin{equation}
P_{\rm
V}(\Delta)d\Delta=A\exp{\left[-\frac{(\Delta^{-2/3}-C_0)^2}{2(2\delta_0/3)^2}
\right]}\Delta^{-\beta}d\Delta,
\end{equation}
with $\delta_0=7.61/(1+z)$ and $\beta=2.23$, 2.35 and 2.48, and $C_0=0.558$,
0.599 and 0.611 at $z=2$, 3 and 4. At $z=6$ they assume $\beta=2.5$, which
corresponds to the distribution of densities of an isothermal sphere, and
solve for $A$ and $C_o$ by requiring the mass and volume to be normalized
to unity. We repeat this procedure to find $P_{\rm V}(\Delta)$ at higher
redshifts. The proportionality of $\delta_0$ to the scale factor is expected
for the growth of structure in an $\Omega_{\rm m}=1$ universe or at high redshift
otherwise, and its amplitude should depend on the amplitude of the power-spectrum. 
The simulations on which the distribution in Miralda-Escude et al.~(2000)
was based assumed $\Omega_{m}=0.4$ in matter, $\Omega_\Lambda=0.6$ in a cosmological 
constant and $\sigma_8=0.79$, close to the values used in this paper.

Equation~(\ref{postoverlap}) provides a good description of the
evolution of the ionization fraction following the overlap of individual ionized 
bubbles, because the ionization fronts are exposed to the mean ionizing radiation
field. However prior to overlap, the prescription is inadequate, due to the large 
fluctuations in
the intensity of the ionizing radiation. A more accurate model to describe
the evolution of the ionized volume prior to overlap was suggested by 
Miralda-Escude et al.~(2000). In our notation the appropriate equation is
\begin{equation}
\frac{d[Q_{\rm i}F_{\rm M}(\Delta_{\rm crit})]}{dz} =
\frac{1}{n^0}\frac{dn_{\gamma}}{dz} - \alpha_{\rm B}(1+z)^3R(\Delta_{\rm
crit})n_{\rm e}Q_{\rm i}\frac{dt}{dz}.
\end{equation} 
or
\begin{equation}
\label{preoverlap}
\frac{dQ_{\rm i}}{dz} = \frac{1}{n^0 F_{\rm M}(\Delta_{\rm
crit})}\frac{dn_{\gamma}}{dz} - \left[\alpha_{\rm B}(1+z)^3R(\Delta_{\rm
crit})n_{\rm e}\frac{dt}{dz} + \frac{dF_{\rm M}(\Delta_{\rm
crit})}{dz}\right]\frac{Q_{\rm i}}{F_{\rm M}(\Delta_{\rm crit})}.
\end{equation} 
In this expression, $Q_{\rm i}$ is redefined to be the volume filling
factor within which all matter at densities below $\Delta_{\rm crit}$ has been
ionized. Within this formalism, the epoch of overlap is precisely defined
as the time when $Q_{\rm i}$ reaches unity. However, we have only a single
equation to describe the evolution of two independent quantities $Q_{\rm
i}$ and $F_{\rm M}$. The relative growth of these depends on the luminosity
function and spatial distribution of the sources. The appropriate value of
$\Delta_{\rm crit}$ is set by the mean free path of the ionizing photons. More
numerous sources can attain overlap for smaller values of $\Delta_{\rm
crit}$. Assuming $\Delta_{\rm crit}$ to be constant with redshift, we find that
results do not vary much (less than 10\% in the optical depth to electron
scattering) for values of $\Delta_{\rm crit}$ ranging from a few to a few
tens. At high redshift, these $\Delta_{\rm crit}$ correspond to mean free paths
comparable to the typical separations between galaxies or quasars. We
assume $\Delta_{\rm crit}=20$ (which lies between the values for galaxies and 
quasars) throughout the remainder of this paper.

\section{The Reionization of Hydrogen and Helium}

Next we describe the joint evolution of the filling factors and ionized
mass fractions for hydrogen and helium using generalizations of
equations~(\ref{postoverlap}) and (\ref{preoverlap}). Let $Q_{\rm H^+}$,
$Q_{\rm He^+}$ and $Q_{\rm He^{\rm ++}}$ be the filling factors of ionized
hydrogen, singly ionized helium and doubly-ionized helium, and $F_{\rm
H^+}$, $F_{\rm He^+}$ and $F_{\rm He^{\rm ++}}$ be the mass-fractions of
ionized hydrogen, singly ionized helium and doubly ionized helium, within the
volumes $Q_{\rm H^+}$, $Q_{\rm He^+}$ and $Q_{\rm He^{\rm ++}}$ respectively. 
For later use we also define\footnote{Note that $Q^{\rm m}$ is the mass-filling factor. The 
mass filling factors for H$^+$ and He$^{++}$ equal $Q_{\rm H^+}F_{\rm H^+}$ and
$Q_{\rm He^{++}}F_{\rm He^{++}}$ respectively. Prior to the overlap of 
HeIII regions, $Q^{\rm m}_{\rm He^+}=Q_{\rm He^{+}}F_{\rm He^{+}}$. However
following the overlap of HeIII regions, $Q^{\rm m}_{\rm He^+}=F_{\rm He^{+}}$.}
the fractions of ionized mass in the universe $Q^{\rm m}_{\rm H^{+}}$, 
$Q^{\rm m}_{\rm He^{+}}$ and $Q^{\rm m}_{\rm He^{++}}$. 

We assign the comoving densities of hydrogen 
and helium; $n_{\rm
H}^{\rm 0}=1.88\times10^{\rm -7}({\Omega_{\rm b}}h^2/{0.022})~{\rm cm^{\rm
-3}}$ and $n_{\rm He}^{\rm 0}=0.19\times10^{\rm -7}({\Omega_{\rm
b}h^2}/{0.022})~{\rm cm}^{\rm -3}$, respectively, and adopt the case B recombination
coefficients\footnote{At high redshift the absorbtion mean-free-path divided by the
speed of light is much smaller than the Hubble time. The appropriate recombination 
coefficient is therefore case B.} for $\rm H^+$, $\rm He^{\rm +}$ and $\rm He^{\rm ++}$ at a
temperature of $10^4$K (Osterbrock~1974; Barkana \& Loeb~2001);
$\alpha_{\rm B}^{\rm H^+}=2.6\times10^{\rm -13}{\rm cm^3~s^{\rm -1}}$,
$\alpha_{\rm B}^{\rm He^{+}}=2.73\times10^{-13}{\rm cm^3~s^{-1}}$ and
$\alpha_{\rm B}^{\rm He^{++}}=13\times10^{-13}~{\rm cm^3~s^{-1}}$. The ionizing
radiation field is described by the comoving densities of ionizing photons for
ionized hydrogen, singly ionized helium and doubly ionized helium, namely
$n_{\rm \gamma}^{\rm H^{\rm +}}$, $n_{\rm \gamma}^{\rm He^{\rm +}}$ and
$n_{\rm \gamma}^{\rm He^{\rm ++}}$, respectively. These quantities
are computed through integration of the corresponding spectra in the 
frequency intervals $3.29\times10^{15}<\nu<5.94\times10^{15}$Hz for HI,
$5.94\times10^{15}<\nu<1.31\times10^{16}$Hz for HeI (He$^{+}$), and
$\nu>1.31\times10^{16}$Hz for HeII (He$^{\rm ++}$). Two classes of sources,
quasars and stars, contribute to the ionizing radiation field. We discuss 
these in turn in the following subsections.

\subsection{Ionizing Photons from Quasars}

To calculate the comoving density of ionizing photons emitted by quasars, 
$n_{\rm \gamma}$, we integrate over both the quasar luminosity function and 
over the mean quasar spectrum (specified in Schirber \& Bullock~2002). Given
the B-band quasar luminosity function $\Phi(L_{\rm B},z)$ at redshift $z$ in units 
of ${\rm Mpc}^{-3}L_{\rm B}^{-1}$, we find for example
\begin{equation}
\frac{dn_{\rm \gamma}^{\rm H^{\rm +}}}{dz} = -\frac{dt}{dz}\int_{\rm
0}^{\infty}dL_{\rm B}\Phi(L_{\rm B},z)\int_{\rm 3.29\times10^{15}{\rm
Hz}}^{5.94\times10^{15}{\rm Hz}}d\nu\frac{L_{\rm \nu}(L_{\rm B})}{h_{\rm
p}\nu},
\end{equation}
where $h_{\rm p}$ is the Planck constant, and $L_{\rm \nu}(L_{\rm B})$ is
the luminosity in units of ${\rm erg~s^{-1}~Hz^{-1}}$ of a quasar with a
B-band luminosity $L_{\rm B}$. At
redshifts below $z=2.5$ we use the luminosity function of Boyle et
al.~(2000). At $z>2.5$ we use the theoretical luminosity function of Wyithe
\& Loeb~(2002); the reader is referred to the original paper for more
details.  This quasar evolution model successfully describes all known
properties of the high redshift quasar luminosity function (Fan et
al.~2001a,b), and reproduces measurements of the black-hole/dark matter
halo mass relation (Ferrarese~2002) as well as estimates of the quasar duty
cycle (Steidel et al.~2002). Since quasars at high redshift appear similar
to typical examples at low redshift (Fan et al.~2001b), the luminosity function 
fully determines the ionizing radiation field. We therefore have almost no 
freedom in our calculation of the ionizing radiation field from quasars at high
redshifts. This allows us to explore the effect of different stellar
radiation fields on the reionization history of the universe.

\subsection{Ionizing Photons from Stars}
\label{stars}

To compute the comoving densities of stellar ionizing photons, ${n_{\rm \gamma}}$,
we integrate over model spectra for stellar populations in the frequency ranges
quoted above to find (for each species) the total number of ionizing photons,
$N_{\rm \gamma}$, emitted per baryon incorporated into stars. We find
\begin{equation}
\label{star_ionize}
\frac{dn_{\rm \gamma}^{\rm H^{\rm +}}}{dz} = N_{\rm \gamma}f_{\rm
esc}f_{\star}\frac{dF_{\rm b}}{dz}n_{\rm b},
\end{equation}
where $n_{\rm b}$ is the local number density of baryons, $f_{\rm esc}$ is the escape 
fraction for the ionizing radiation, $f_{\star}$ is the fraction of baryons within 
a galaxy that are incorporated in stars, and $F_{\rm b}(z)$ is the fraction of 
baryons in the universe at redshift $z$ that collapsed and cooled inside galaxies. 
To compute $F_{\rm b}(z)$
we require the dark matter collapse fraction in halos $F_{\rm col}(z, T_{\rm min})$ 
with virial temperatures larger than $T_{\rm min}$, given by
\begin{equation}
F_{\rm col}(z, T_{\rm min})=\frac{1}{\rho_{\rm m}}\int_{M_{\rm min}(T_{\rm min})}^\infty dM 
M\frac{dn_{\rm ps}}{dM},
\end{equation}
where $\rho_m$ is the co-moving mass density in the universe, and $dn_{\rm ps}/dM$ is
the Press-Schechter~(1974) mass function for dark matter halos.
In regions of the universe that are neutral, the critical virial temperature (the 
temperature at which atomic cooling becomes efficient) is 
$T_{\rm min}=10^4$K (a circular velocity of $v_{\rm cir}\sim10$km$\,$s$^{-1}$). 
In reionized regions, infall of gas is suppressed below a virial temperature of 
$T_{\rm min}\sim2.5\times10^5$K ($v_{\rm cir}\sim50$km$\,$s$^{-1}$, Thoul \& Weinberg~1996). Given $F_{\rm col}(z,10^4K)$
and $F_{\rm col}(z,2.5\times10^5K)$ we can take the baryonic 
collapse fraction
\begin{equation}
\label{fb}
F_{\rm b}(z)=Q^{\rm m}_{\rm H^+}F_{\rm col}(z, 2.5\times10^5\mbox{K}) + (1-Q^{\rm m}_{\rm H^+})F_{\rm col}(z, 10^4\mbox{K}).
\end{equation}
This prescription for calculating the stellar
ionizing field will be referred to hereafter as case A.  Note that $f_{\rm
esc}$ could be different for hydrogen and helium; however, we adopt a
single value for it assuming that this value is dictated by the geometry of
the optically-thick gas that allows limited escape routes which are common
for all ionizing photons (Wood \& Loeb 2000 and references therein).

We also consider a second evolution for the stellar ionizing field,
denoted hereafter as case B. A recent study (Kauffmann et al.~2002)
has shown that in a large sample of local galaxies, the ratio
$\epsilon=M_\star/M_{\rm halo}$ (where $M_\star$ and $M_{\rm halo}$
are the total stellar and dark matter halo masses respectively) scales
as $\epsilon\propto M_{\rm halo}^{2/3}$ for
$M_{\star}<3\times10^{10}M_{\odot}$, but is constant for larger
stellar masses. Note that the star-formation efficiency is
proportional to $\epsilon$.  Since star-formation is thought to be
regulated by supernova feedback (Dekel \& Silk 1986), the important
quantity is the depth of the galactic potential well, or equivalently
the halo circular velocity.  Using the stellar mass Tully-Fisher
relation of Bell \& De Jong~(2001), we find the threshold circular
velocity $v_\star=176~{\rm km~s^{-1}}$ that at $z=0$ corresponds to a
stellar mass of $3\times10^{10}M_{\odot}$. In this case we define
$f_\star$ as the star-formation efficiency in galaxies with circular
velocities larger than $v_\star$, and calculate the effective product
of the dark matter collapse fraction and star-formation efficiency
\begin{equation}
f_\star F_{\rm col}(z)=\frac{1}{\rho_{\rm m}}\int_{M_{\rm min}(T_{\rm
min})}^\infty dM \epsilon f_\star M\frac{dn_{\rm ps}}{dM},
\end{equation}
where $\epsilon=1$ for $M>M_{\rm halo}^\star$ and
$\epsilon=\left({M}/{M_{\rm halo}^\star}\right)^{2/3}$ for $M<M_{\rm
halo}^\star$. The value of $M_{\rm halo}^\star(z)$ is calculated from
\begin{equation}
M_{\rm halo}^\star(z)=4.3\times10^{10}h^{-1}\left(\frac{v_\star}{{\rm 176
km~s^{-1}}}\right)^3 \left[\frac{\Omega_{\rm
m}}{\Omega_{\rm m}^{\rm z}}\frac{\Delta_{\rm crit}}{18\pi^2}
\right]^{-1/2}\left(\frac{1+z}{10}\right)^{-3/2}M_{\odot},
\end{equation}
where $\Delta_{\rm crit}=18\pi^2+82d-39d^2$, $d=\Omega_{\rm m}^{\rm
z}-1$ and $\Omega_{\rm m}^{\rm z}={\Omega_{\rm
m}(1+z)^3}/[{\Omega_{\rm m}(1+z)^3+\Omega_\Lambda}]$. Finally, the
value of $f_\star F_{\rm b}$ required for calculation of $dn_{\rm
\gamma}/dz$ may then be computed in analogy with case A, and
substituted into equation~(\ref{star_ionize}).

In both cases A and B we make the distinction between the ionizing
radiation field due to a possible early population of zero-metallicity
stars, and the metal enriched stars observed at lower redshifts. It is
thought that the primordial initial mass function favored massive
stars (Bromm, Copi \& Larson~1999, 2001; Abel, Bryan \& Norman~2000;
Mackey, Bromm, Hernquist~2002).  The possible existence of this
population is very important for reionization because the spectrum of
these stars would result in an order of magnitude more ionizing
photons per baryon incorporated into stars (Bromm, Kudritzki, \& Loeb
2001). The formation of the very massive stars was suppressed as the
material out of which stars form was enriched with metals.  The
fraction of the ionizing photons produced by metal-free stars depends
on several unknown parameters, including the mixing efficiency of
metals, the environments in which new stars form, and most
importantly, the threshold metallicity above which star formation is
transformed from being dominated by massive stars to a Scalo~(1998)
initial mass function (IMF). The threshold metallicity is believed to
be small; Bromm et al.~(2001) argue for a threshold $\frac{Z_{\rm
thresh}}{Z_\odot}\la 10^{-3}$ of the solar metallicity value.  The
efficiency of mixing of metal enriched outflows from star forming
galaxies to the surrounding IGM is even more uncertain; Scannapieco,
Ferrara \& Madau~(2002) find that the mass weighted mean metallicity
can reach values greater than $10^{-3}$ of the solar value at
redshifts as high as 20, and note that the average metallicity scales
with star formation efficiency, supernovae rate, and the fraction of
supernovae energy that is channeled into outflows.  The average
metallicity increases roughly exponentially with redshift as it is
modulated by the exponential growth in the collapse fraction of
baryons at high redshifts.

Since the first metals were produced by supernovae, it is reasonable
to suppose that the enrichment of the IGM with metals at redshift $z$
is proportional to the mass of baryons that has formed stars by that
redshift. The average metallicity at a redshift $z$ can then be
written as
\begin{equation}
\label{gradualenrich}
\frac{Z(z)}{Z_\odot}=C_{\rm metal}\frac{Z_{\rm thresh}}{Z_\odot}\int_\infty^z dz \frac{dF_{\rm b}}{dz},
\end{equation}
where $C_{\rm metal}$ is a constant. Note that the baryonic collapse
fraction $F_{\rm b}$, which is computed from equation~(\ref{fb}) is
affected by the reionization history.  The formation of metal-rich
stars is assumed to be proportional to the average metalicity, with
the fraction reaching unity when the average metalicity of the IGM
reaches $\frac{Z_{\rm thresh}}{Z_\odot}$. Before
$\frac{Z(z)}{Z_\odot}$ reaches \ $\frac{Z_{\rm thresh}}{Z_\odot}$ we
may approximate the fraction of star formation at redshift $z$ that is
in zero-metallicity stars to be $(1-\frac{Z(z)}{Z_\odot}/\frac{Z_{\rm
thresh}}{Z_\odot})$, while after $\frac{Z(z)}{Z_\odot}$ reaches
$\frac{Z_{\rm thresh}}{Z_\odot}$ we may assume all star formation to
have a Scalo~(1998) IMF\footnote{The situation in reality is more
complicated since mixing of metals is incomplete and the formation
sites of new stars are correlated with the enriched regions.  Our
discussion uses the minimum number of free parameters to describe this
complicated process.}.

For a given star formation efficiency and escape fraction, the above
prescription yields a characteristic redshift for the build-up of the
threshold metal enrichment.  The exponential growth in the collapse
fraction with redshift implies that the evolution in the mode of star
formation resembles a step-function.  We therefore define a transition
redshift $z_{\rm tran}$ below which metal-rich stars with a
Scalo~(1998) IMF dominate the production rate of ionizing photons.

Through most of the paper we present results in terms of $z_{\rm
tran}$, but return to a justification of this choice in
\S~\ref{justification}.  At redshifts above $z_{\rm tran}$, we use
values of $N_{\rm \gamma}$ calculated for each of the three species
$\rm H^+$, $\rm He^+$ and $\rm He^{++}$ assuming massive ($\ga 100
M_\odot$) zero-metallicity stars and the generic spectrum calculated
by Bromm, Kudritzki \& Loeb~(2001). The resulting values of $N_{\rm
\gamma}$ are $\sim14020$ for H$^+$, $\sim 25200$ for $\rm He^+$ and
$\sim 4480$ for $\rm He^{++}$. At lower redshifts we assume metal
enriched stars (1/20th solar metallicity) with a Scalo~(1998)
mass-function, and use spectral information from the stellar
population model of Leitherer et al.~(1999)\footnote{Model spectra of
star-forming galaxies were obtained from
http://www.stsci.edu/science/starburst99/.}.  This results in values
of $N_{\rm \gamma}\sim3250$ for H$^+$, $N_{\rm \gamma}\sim 1020$ for
$\rm He^+$ and $N_{\rm \gamma}\sim 0.2$ for $\rm He^{++}$.

\subsection{Constraints on the Evolution of $\rm H^+$, 
$\rm He^+$ and $\rm He^{++}$ Ionization Fronts}

%FIGURE 1
\begin{figure*}[htbp]
\epsscale{1}
\plotone{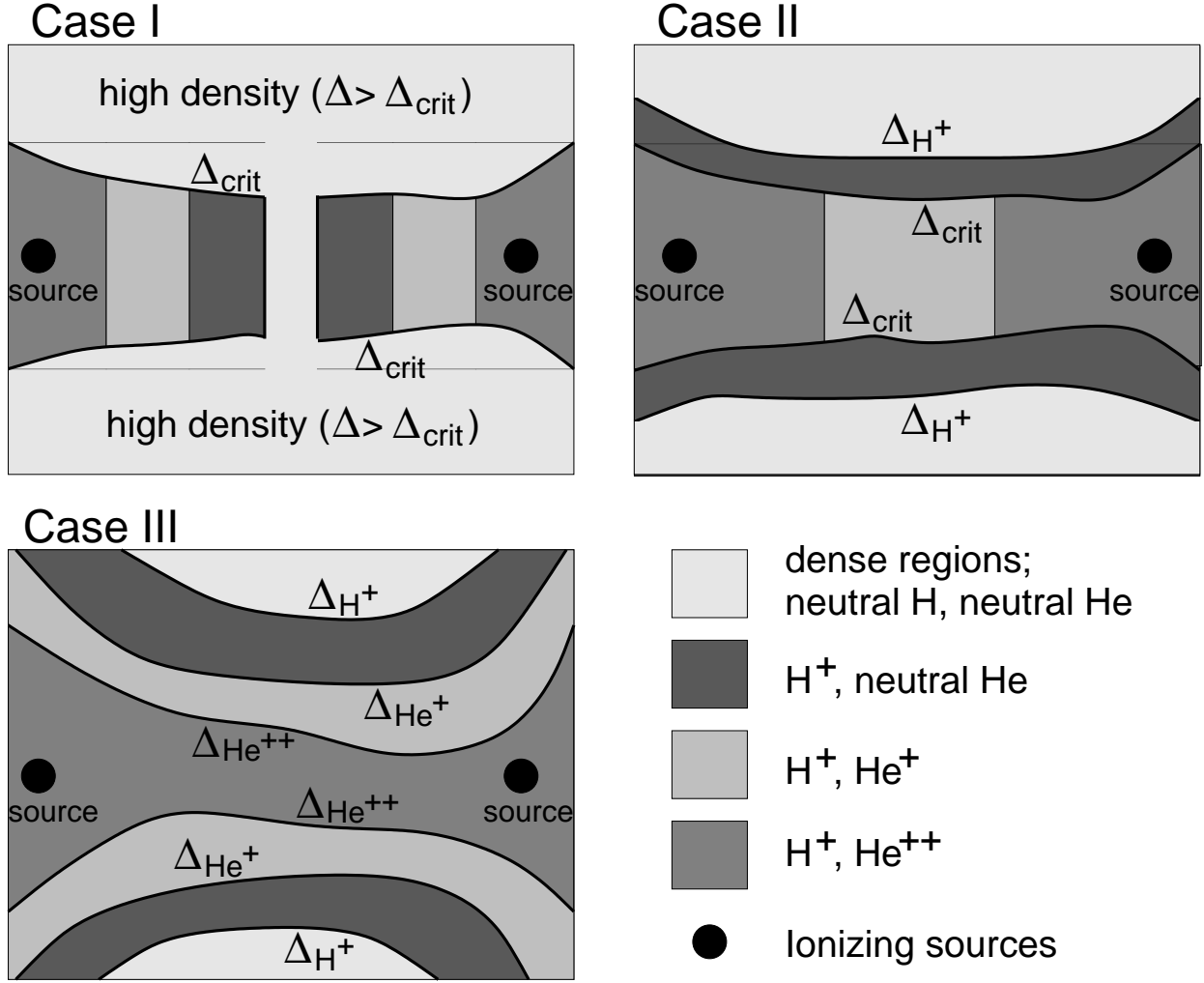}
\caption{\label{fig1} Schematic diagram of the process of
reionization. Prior to the overlap of hydrogen, reionization proceeds
through low density regions (below $\Delta_{\rm crit}$) for both hydrogen
and helium. Following the overlap of HII regions, the hydrogen reionization
front moves into denser regions but helium reionization continues to
proceed through regions with density below $\Delta_{\rm crit}$. Finally,
after the overlap of HeIII regions, the reionization of helium is free
to proceed into denser parts of the IGM.}
\end{figure*}

The typical distance that an ionizing photon of neutral helium can travel
through neutral hydrogen having a cross-section $\sigma_{\rm H}(\nu)$ before the
optical depth reaches unity is the absorption mean-free-path $\lambda$,
where
\begin{equation}
\lambda^{-1} = \frac{\int_{5.94\times10^{15}{\rm
Hz}}^{1.31\times10^{16}{\rm Hz}}~~d\nu (L_{\nu}/\nu)\sigma_{\rm H}(\nu)
n_{H}^{0}(1+z)^3 }{\int_{5.94\times10^{15}{\rm Hz}}^{1.31\times10^{16}{\rm
Hz}}~~d\nu L_{\nu}/\nu}.
\end{equation}
We find $\lambda \sim 4\left[\left(1+z\right)/10\right]^{-3}$kpc. This distance is
much shorter than $\sim1$Mpc which is the typical size of an HII region
around a $10^8M_{\odot}$ halo prior to reionization (Barkana \&
Loeb~2000). Therefore, at $z\sim 10$ the $\rm He^+$ propagation front can
only lead the $\rm H^+$ propagation front by a small fraction of the size of the HII
region, and the excess $\rm He^{+}$ ionizing photons photo-ionize
additional hydrogen. On the other hand, most ionizing photons for hydrogen
($\nu>5.94\times10^{15}$Hz) are not absorbed by helium, and so the
expansion of the HII region is not inhibited by the presence of
helium. Obviously, due to the absence of singly ionized helium the $\rm
He^{++}$ front cannot propagate beyond the $\rm H^{+}$ front.

Following these considerations, we impose two restrictions on the evolution
of $Q_{\rm He^+}$ and $Q_{\rm He^{++}}$. First, $Q_{\rm He^{++}}\le Q_{\rm
H^{+}}$. Second, $Q_{\rm He^{+}}\le Q_{\rm H^{+}}-Q_{\rm
He^{++}}$. Following helium overlap, we impose the same restrictions on the
evolution of the mass fraction, i.e. $F_{\rm He^{++}}\le F_{\rm H^{+}}$ and
$F_{\rm He^{+}}\le F_{\rm H^{+}}-F_{\rm He^{++}}$. The full equations
governing the evolution of the filling factors and mass fractions are
presented in the Appendix. The equations are split into three cases
describing three different epochs. The evolution of the filling factors
prior to hydrogen overlap are described in case~I. 
Case~II describes the evolution of the filling factors and of the ionized
hydrogen mass fraction following the overlap of HII regions, but prior to
the overlap of HeII regions. Finally, case III describes the evolution of
the ionized mass-fractions following the overlap of cosmological HeII
regions. Figure~\ref{fig1} provides a schematic representation of the three epochs 
in reionization history.
  
\section{Reionization Histories}
\label{histories}

%FIGURE 2
\begin{figure*}[htbp]
\epsscale{1}
\plotone{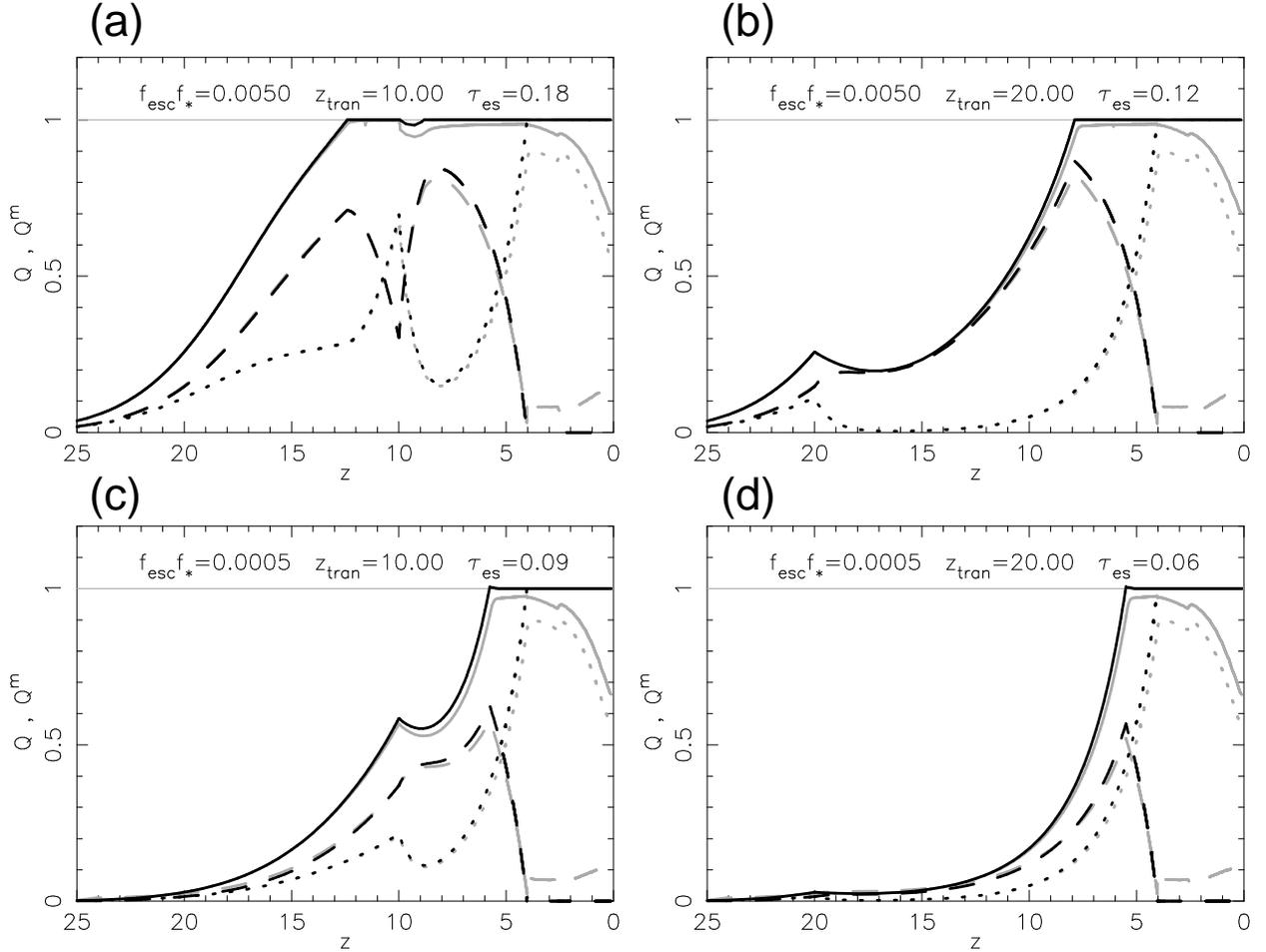}
\caption{\label{fig2} Sample reionization histories assuming case A star
formation. Dark lines represent the evolution of the volume filling factors
$Q_{\rm H^+}$, $Q_{\rm He^+}$ and $Q_{\rm He^{++}}$, while the light lines show 
the evolution of the fractions of ionized mass in the universe $Q^{\rm m}_{\rm H^+}$, 
$Q^{\rm m}_{\rm He^+}$ and $Q^{\rm m}_{\rm He^{++}}$.
The solid, dashed and dotted lines refer to ${\rm H^+}$, $\rm
He^+$ and $\rm He^{++}$, respectively.}
\end{figure*}

%FIGURE 3
\begin{figure*}[htbp]
\epsscale{1}
\plotone{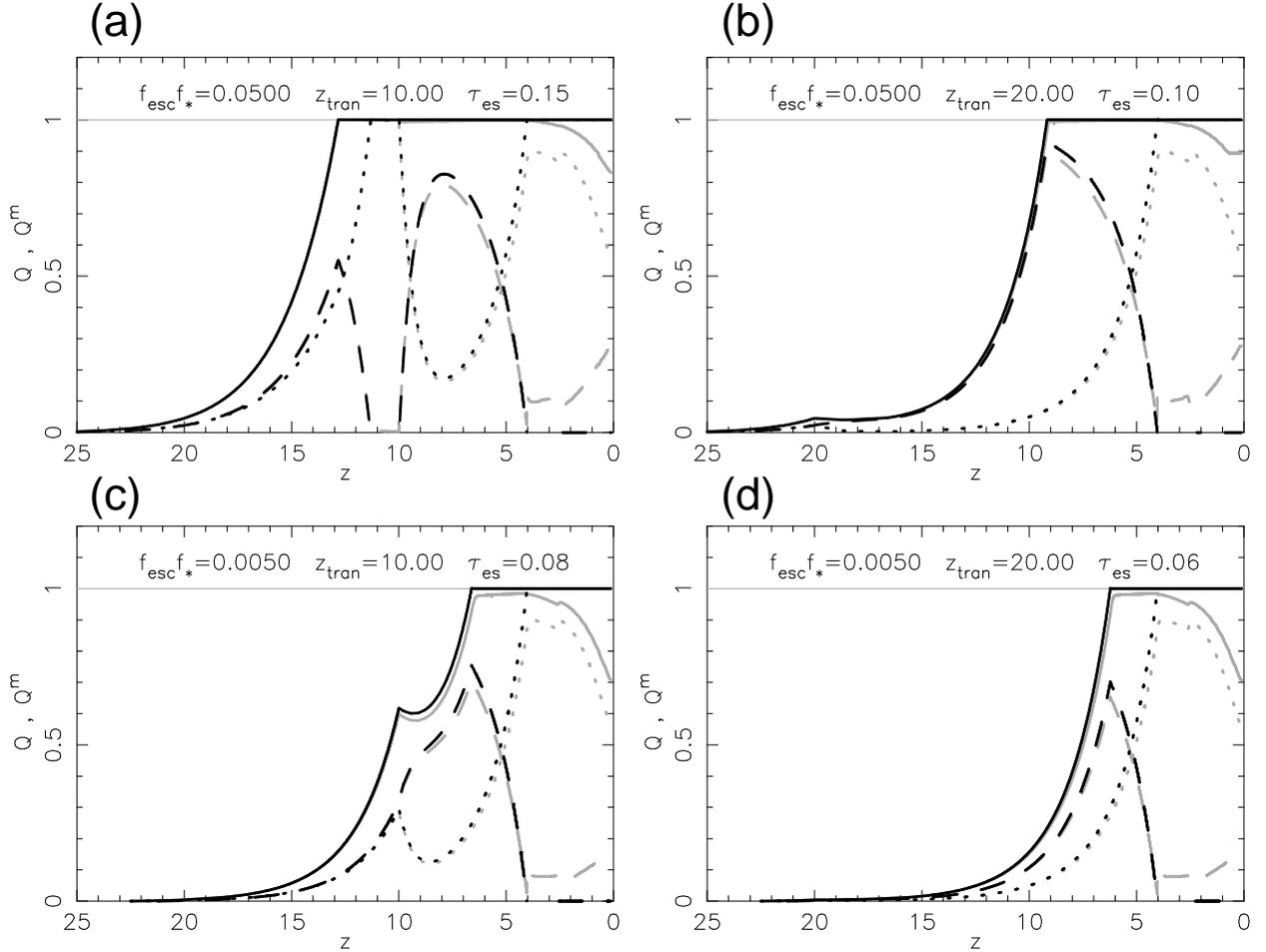}
\caption{\label{fig3} Sample reionization histories assuming case B star
formation. Dark lines represent the evolution of the volume filling factors
$Q_{\rm H^+}$, $Q_{\rm He^+}$ and $Q_{\rm He^{++}}$, while the light lines show 
the evolution of the fractions of ionized mass in the universe $Q^{\rm m}_{\rm H^+}$, 
$Q^{\rm m}_{\rm He^+}$ and $Q^{\rm m}_{\rm He^{++}}$.
The solid, dashed and dotted lines refer to ${\rm H^+}$, $\rm
He^+$ and $\rm He^{++}$, respectively.}
\end{figure*}
In this section we derive reionization histories for different values
of $z_{\rm tran}$ and $f_{\rm esc}f_\star$. These histories are shown in
figures~\ref{fig2} (for case A star-formation) and \ref{fig3} 
(for case B star-formation).

For case A we show results for combinations of $f_{\rm
esc}f_{\star}=0.0005$ and 0.005, and $z_{\rm tran}=10.0$ and 20.0. We note
the following features. Large values of $f_{\rm esc}f_{\star}$ allow the
massive zero metallicity stars to reionize both hydrogen and helium in the
universe at very early times. If the zero metallicity stars do not reionize
helium, quasars result in the overlap of HeII regions by $z\sim4$,
consistent with observations showing transmission just blueward of the
helium Ly$\alpha$ line at $z\sim3$ (Jacobsen et al 1994; Tytler 1995;
Davidsen et al. 1996; Hogan et al. 1997; Reimers et al. 1997; Heap et
al. 2000; Kriss et al. 2001; Smette et al. 2002) as well as evidence for a
temperature rise in the IGM at $z\sim3.4$ indicating the reionization of
helium at that time (Schaye et al. 2000; Theuns et al. 2002).
Interestingly, we also find that it is possible for hydrogen and/or helium
to have been reionized twice. This situation arises if the zero-metallicity
stars reionize the universe prior to $z_{\rm tran}$, but below $z_{\rm
tran}$ the recombination rate is sufficiently high that the softer spectra
of normal stars cannot maintain overlap. Overlap of HII regions is then
re-achieved at a later time as the collapse fraction grows and the density
of the IGM is lowered. A second overlap of HeII regions results from the
rise in quasar activity at $z\la10$. Similar results are found in case B,
for values of $f_{\rm esc}f_{\star}$ that are a factor of $\sim10$ larger
than in case A.  

We would like to reiterate the point that multiple reionization epochs
occur for some sets of parameters as a result of multiple peaks in the
emissivity of ionizing photons as a function of cosmic time. In such cases
the first emissivity peak occurs at $z_{\rm tran}$, at which point it
drops due to the softer spectra of metal enriched stars. Later, when the
collapsed fraction of baryons has reached a sufficiently high value, the
resulting increase in emissivity further reionizes the IGM. As we will
show in \S~\ref{justification}, this double peaked behavior is still seen
if a smoothly varying fraction of collapsed objects at metallicities below
the threshold produce the hard spectra rather than sources emitting prior
to a fixed $z_{\rm tran}$. A double peaked emissivity will always arise so
long as the relevant metal enrichment time is less than the time required
for the collapse fraction to change by a factor equal to the ratio between
the emissivities of the first generation of stars and subsequent metal
enriched stars.\footnote{Note that even if the mixing of metals is partial
and the intergalactic filling factor of metals is smaller than unity
(e.g., Scannapieco et al. 2002), new stars are likely to form in regions
that were already enriched because galaxies tend to be clustered on large
scale sheets and filaments.  This effect needs to be included when
evaluating the effective $z_{\rm tran}$ from numerical simulations.}

There are two small, sharp jumps in the reionization histories that are
artifacts, but warrant explanation. The first is in the evolution of 
$Q^{\rm m}_{\rm He^+}$ immediately following the overlap of HeIII regions.
This small jump (note that the curve is smooth, not discontinuous) arises 
because our formalism demands $Q^{\rm m}_{\rm He^+}=0$
at this epoch. However following the overlap of He$^{++}$, the He$^{+}$ front is 
free to rapidly expand, powered by the strong ionizing radiation field, and the 
mass fraction $F_{\rm He^{+}}$ therefore grows until the He$^{+}$ front reaches
the H$^+$ front. The second jump occurs at
$z=2.5$ where we switch from computation of the ionizing flux due to quasars
using the model quasar luminosity function, to calculation using the 
empirical luminosity function. Neither is a perfect description around this 
redshift, and there is a discontinuous jump in the luminosity density, and 
hence in the ionizing radiation density, resulting in a jump in the reionization 
histories. 

At early times the rise in temperature associated with reionization curtails
star-formation. The effect of this is clearly seen in panel (a) of figure~\ref{fig2}.
Here the zero metallicity stars rapidly increase the ionized fraction, but as  
$Q_{\rm H^+}$ approaches unity, the ionizing flux is reduced, and the rate of
reionization slows. This feature is not seen in cases where overlap occurs at lower 
redshift (e.g. panel (d) in figure~\ref{fig2}).
Many combinations of the parameters $z_{\rm tran}$ and $f_{\rm esc}f_\star$
predict significant reionization without overlap due to the early presence of 
metal-free stars, resulting in a peak in the ionized fraction centered on the 
transition redshift, followed by recombinations before further reionization 
 due to normal stars and quasars. We term these peaks {\it failed overlaps}.
In section \S~\ref{visability} we show that these failed overlaps leave a
significant imprint on the visibility function for electron scattering of CMB 
photons.

\subsection{The Optical Depth to Electron Scattering}

Different reionization histories result in different densities of electrons
as a function of redshift. One simple probe of the reionization history is
the optical depth to electron scattering $\tau_{\rm es}$, which depend on the
mass filling factors $Q^{\rm m}$
\begin{equation}
\tau_{\rm es} = \int_0^{1000}dz\frac{cdt}{dz}\sigma_{T}\left[Q^{\rm m}_{\rm
H^+}n_{\rm H}^{\rm 0} + Q^{\rm m}_{\rm He^+}n_{\rm He}^{\rm
0} + 2Q^{\rm m}_{\rm H^{++}}n_{\rm He}^{\rm 0}\right](1+z)^3,
\end{equation}
where $\sigma_{T}=6.652\times 10^{-25}~{\rm cm^2}$ is the Thomson cross-section.
In the following section we plot contours of $\tau_{\rm es}$ as a function of 
$z_{\rm tran}$ and $f_{\rm esc}f_{\star}$.

\section{Joint Constraints on $z_{\rm tran}$ and $f_{\rm esc}f_{\star}$}

%FIGURE 4
\begin{figure*}[htbp]
\epsscale{1}
\plotone{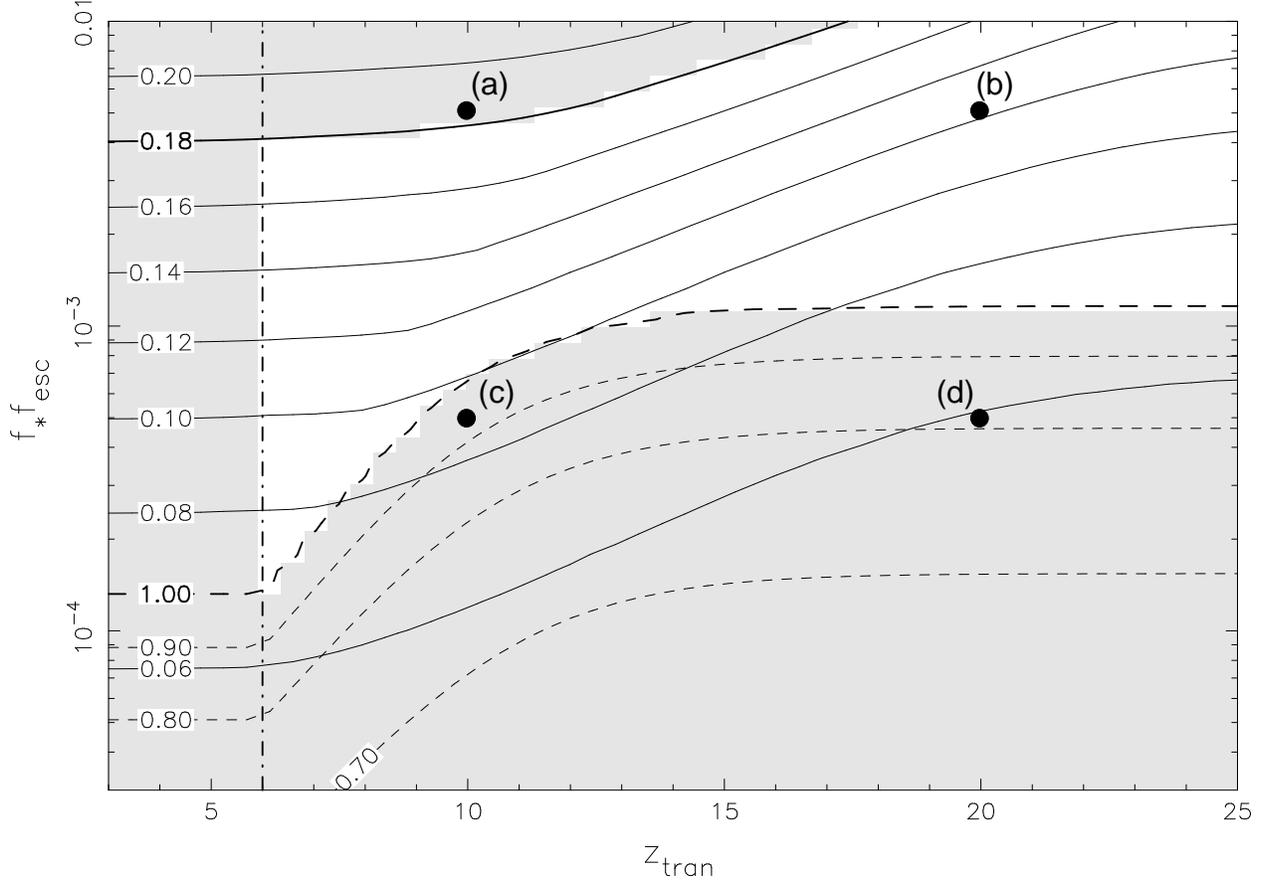}
\caption{\label{fig4} Contours of optical depth for Thomson scattering,
$\tau_{\rm es}$ (solid lines) and contours of $Q_{\rm H^+}$ at $z=6$ (dashed lines), 
as a function of $f_{\rm esc}f_{\star}$ and $z_{\rm tran}$. Case A star-formation
was assumed, and the shaded regions are excluded by existing observations. The
4 labeled points refer to the locations of the parameter sets used in the example 
histories presented in figure~\ref{fig2}.}
\end{figure*}

%FIGURE 5
\begin{figure*}[htbp]
\epsscale{1}
\plotone{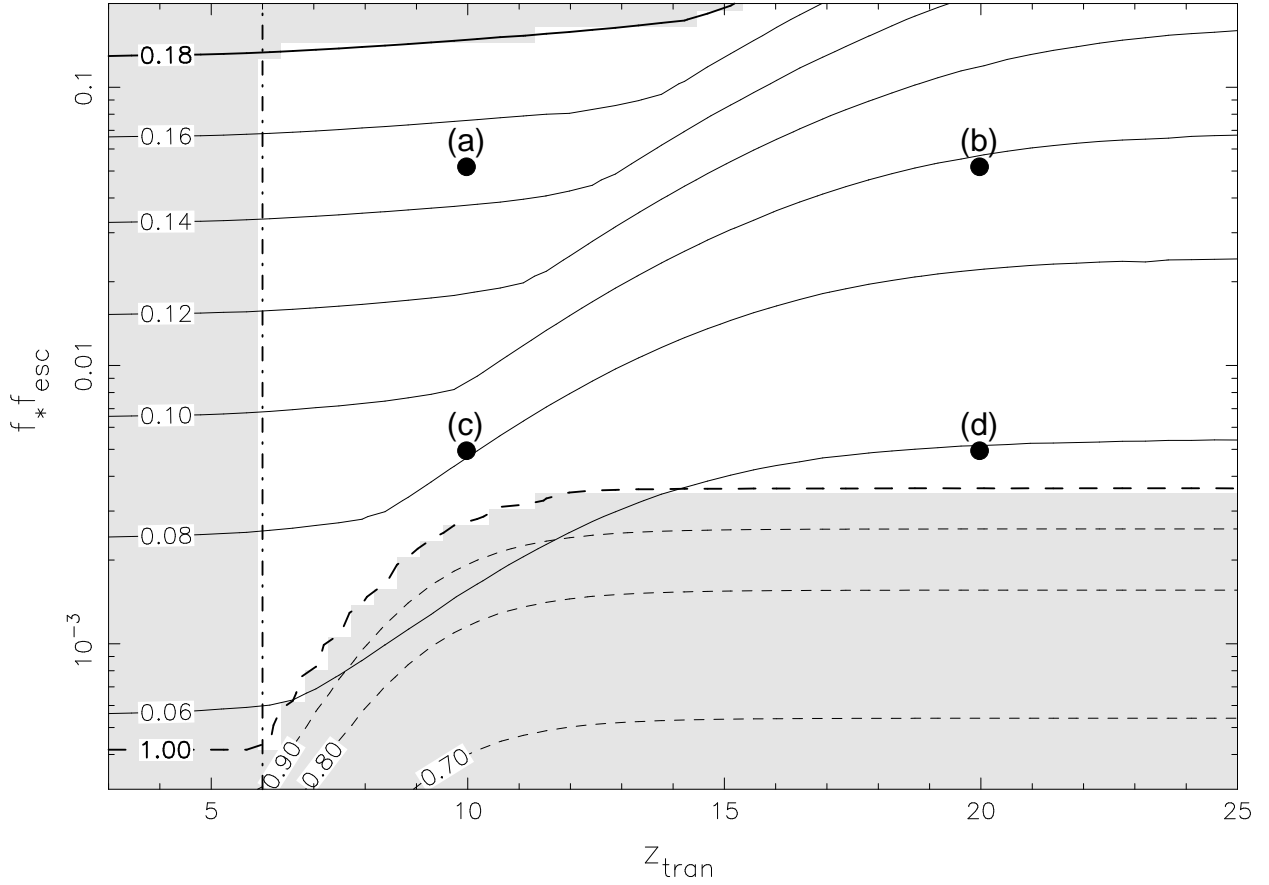}
\caption{\label{fig5} Contours of optical depth for Thomson scattering,
$\tau_{\rm es}$ (solid lines) and contours of $Q_{\rm H^+}$ at $z=6$ (dashed lines), 
as a function of $f_{\rm esc}f_{\star}$ and $z_{\rm tran}$. Case B star-formation
was assumed, and the shaded regions are excluded by existing observations. The
4 labeled points refer to the locations of the parameter sets used in the example 
histories presented in figure~\ref{fig3}.}
\end{figure*}

In this section we discuss the constraints that current observations place
on $z_{\rm tran}$ and $f_{\rm esc}f_\star$.
First we discuss the variation of $\tau_{\rm es}$ with $z_{\rm
tran}$ and $f_{\rm esc}f_{\star}$. The optical depth for electron
scattering is constrained from recent CMB anisotropy experiments to a value
$\tau_{\rm es}<0.18$ (Wang \& Tegmark~2002). An upper limit on $\tau_{\rm
es}$ implies that there is maximum redshift beyond which the universe was
neutral. As a second constraint, we note that the spectra of quasars at 
$z\la 6$ do not show a Gunn-Peterson~(1965) trough, indicating that HII 
regions had achieved overlap earlier than that time.

As we have seen in the previous section, the universe is reionized
earlier for higher values of $f_{\rm esc}f_{\star}$, which result in
higher values of $\tau_{\rm es}$. Hence $f_{\rm esc}f_{\star}$ cannot
be too large so as not to violate the limit $\tau_{\rm es}<0.18$. On
the other hand, if the product $f_{\rm esc}f_{\star}$ is too small,
then the ionizing radiation field will be too weak for the HII regions
to overlap by $z\sim6$.  If the transition redshift $z_{\rm tran}$ is
large, then $f_{\rm esc}f_{\star}$ may obtain higher values since the
harder spectra of metal free stars are only available to the ionizing
radiation field at high redshifts when the IGM is dense and the
collapse fraction is small. However if $z_{\rm tran}$ is low, then
$f_{\rm esc}f_{\star}$ may take smaller values and still result in
overlap of HII regions by $z\sim6$. In addition to the above
constraints, observations suggest that the IGM was metal enriched by
$z\sim6$ (Songaila 2001) to a level that would lead to a Scalo stellar
mass function (Bromm et al. 2001) if new galaxies formed out of it. As
a result, $z_{\rm tran}$ is limited to be larger than six.

Figures~\ref{fig4} and \ref{fig5} show contours of $\tau_{\rm es}$ (solid
lines) as a function of $z_{\rm tran}$ and $f_{\rm esc}f_{\star}$ for
case~A and B star-formation respectively. Also shown are contours of
$Q_{\rm H^+}$ at $z=6$ (dashed lines), and a vertical line at $z_{\rm
tran}=6$ (dot-dashed line). The regions excluded by the aforementioned
constraints are shaded grey. For our case A star-formation model we find
that $f_{\rm esc}f_{\star}\la0.004$ if $z_{\rm tran}\la10$. For a higher
$z_{\rm tran}$, we find that larger values of $f_{\rm esc}f_\star$
are possible. If $z_{\rm tran}\sim6$, $f_{\rm esc}f_{\star}$ may be as low
as $10^{-4}$, though for $z_{\rm tran}\ge15$ we find $f_{\rm
esc}f_{\star}\ga0.001$. We find that the lower limits on $f_{\rm esc}f_\star$
are larger by a factor of $\sim3$ for our case B star-formation model. However since the
intensity of the stellar radiation field is curtailed by feedback at high
redshift in this model, we find that $f_{\rm esc}f_{\star}$ can be as large
as $\sim0.1$ for $z_{\rm tran}\la15$, or even larger for $z_{\rm tran}\ga15$.
Interestingly, the contours for $\tau_{\rm es}$ and $z_{\rm reion}$ are not
parallel. A significant range of $\tau_{\rm es}$ exist for different $z_{\rm tran}$
(particularly where case A star-formation is assumed). 
This indicates that measurement of optical depth does not determine the overlap 
redshift. We discuss the reason for this puzzling result in \S~\ref{visability}.

%FIGURE 6
\begin{figure*}[htbp]
\epsscale{1}
\plotone{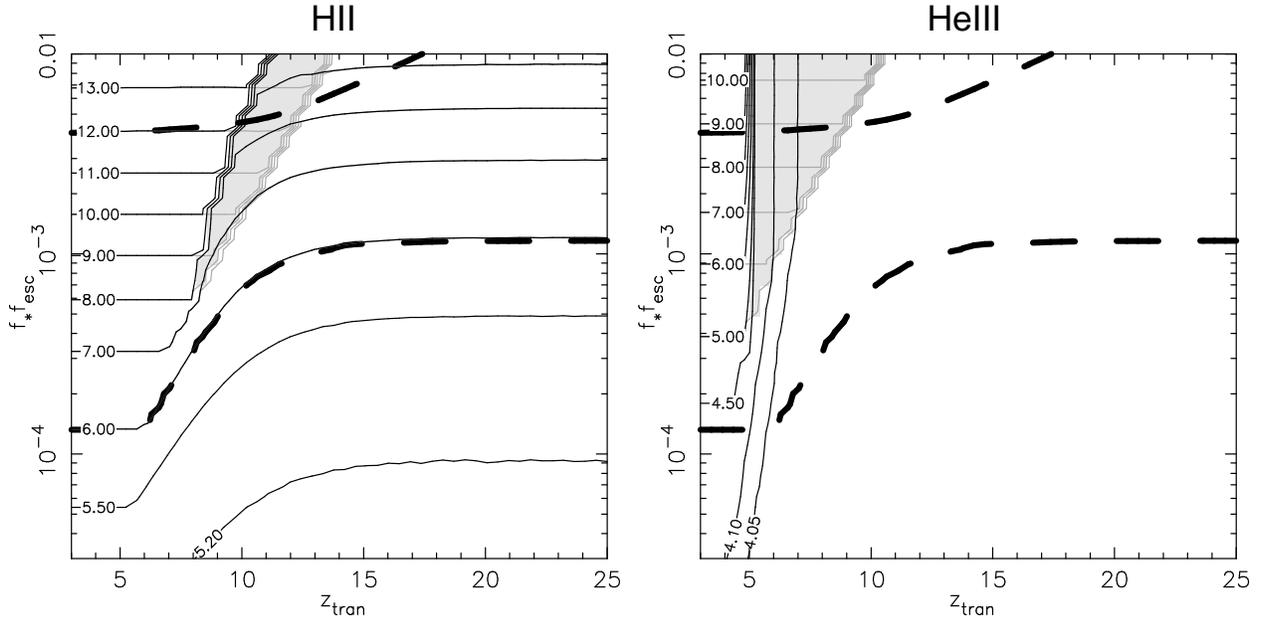}
\caption{\label{fig6} Contours of overlap redshift for hydrogen (left
panel) and fully-ionized helium (right panel), as a function of $f_{\rm
esc}f_{\star}$ and $z_{\rm tran}$ assuming case A star-formation. The dark
contours show the most recent overlap. Parameter pairs that produce two 
overlap epochs lie in the shaded region. The grey contours describe the
redshifts of these earlier overlaps.
Note that the regions below and to the right of the lowest redshift
contour shown have nearly constant overlap redshift. This arises because
quasars dominate the reionization process for these parameter pairs. Also
shown for reference are the contours of $\tau_{\rm es}=0.18$ and $Q_{\rm H^+}=1$
at $z=6$ (thick dashed lines). Parameter pairs not lying between these lines
are excluded by current observation.}
\end{figure*}

%FIGURE 7
\begin{figure*}[htbp]
\epsscale{1}
\plotone{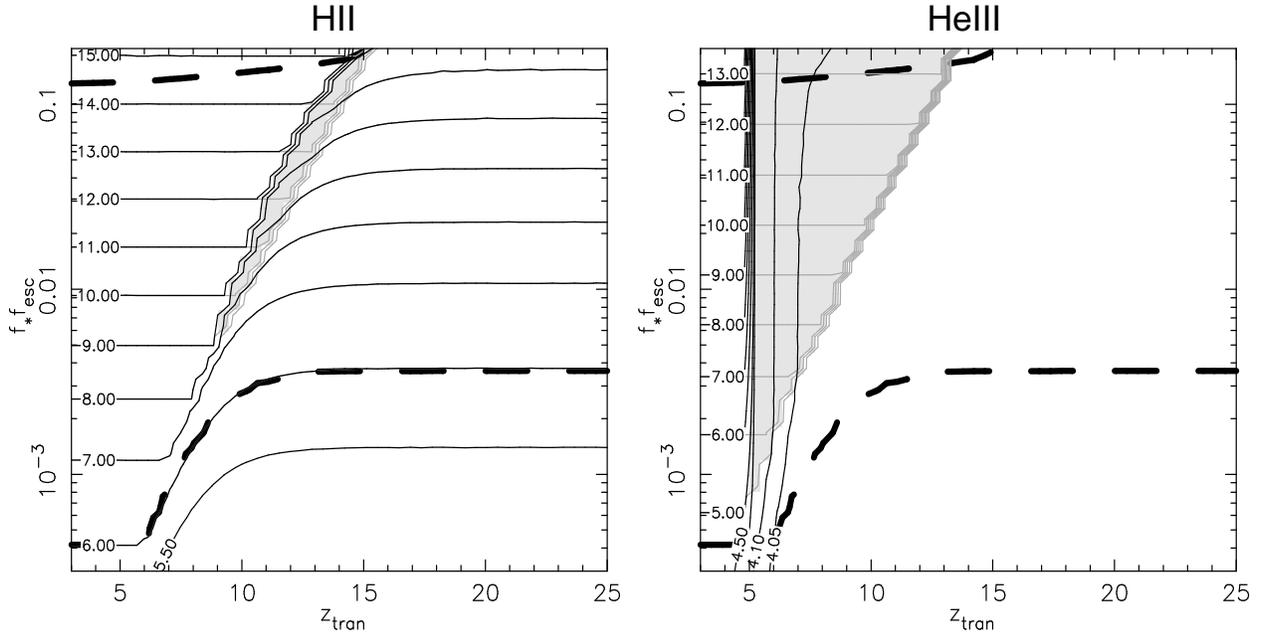}
\caption{\label{fig7} Contours of overlap redshift for hydrogen (left
panel) and fully-ionized helium (right panel), as a function of $f_{\rm
esc}f_{\star}$ and $z_{\rm tran}$ assuming case B star-formation. The dark
contours show the most recent overlap. Parameter pairs that produce two 
overlap epochs lie in the shaded region. The grey contours describe the
redshifts of these earlier overlaps.
Note that the regions below and to the right of the lowest redshift
contour shown have nearly constant overlap redshift. This arises because
quasars dominate the reionization process for these parameter pairs. Also
shown for reference are the contours of $\tau_{\rm es}=0.18$ and $Q_{\rm H^+}=1$
at $z=6$ (thick dashed lines). Parameter pairs not lying between 
these lines are excluded by current observation.}
\end{figure*}

In \S~\ref{histories} we mentioned that hydrogen and/or helium could be reionized twice,
and showed examples of reionization histories that exhibit this behavior. In 
figures~\ref{fig6} and \ref{fig7} we show contours of the overlap redshift for hydrogen
(left panels), and helium (right panels), for case A and case B star-formation models.
The redshift of the most recent overlap is described by the dark contours. At high
values of $z_{\rm tran}$ and low values of $f_{\rm esc}f_\star$, the overlap redshift
is almost independent of star-formation since then quasars dominate the ionizing
radiation field. This is particularly true for helium, which is ionized by quasars 
at $z\sim4$ in the absence of any stellar ionizing radiation. Cases where 
there was an additional earlier overlap are denoted by the grey contours.
The region of the $[z_{\rm tran},f_{\rm esc}f_\star]$ plane that results in multiple 
overlap epochs is shaded grey. We see that this area covers a significant portion of 
parameter space, and therefore that it is quite possible the universe was 
reionized twice. The contours representing the upper limit of $\tau_{\rm es}<0.18$,
and the lower limit for the overlap redshift $z>6$ are also shown (thick dashed 
lines). As mentioned earlier, parameters in the region between these limits are 
not excluded by current observations. 

\section{The Visibility Function}
\label{visability}

The visibility function $g(z)d\eta$ describes the probability that an observed 
photon was scattered between conformal times $\eta(z)$ and
$\eta+\Delta\eta$
\begin{equation}
g(z) = \frac{d\tau_{\rm es}}{d\eta}e^{-\tau_{\rm es}}=\sigma_{\rm T}\left[Q^{\rm m}_{\rm H^+}n_{\rm H}^{\rm 0} + Q^{\rm m}_{\rm He^+}n_{\rm He}^{\rm 0} + 2Q^{\rm m}_{\rm He^{++}}n_{\rm He}^{\rm 0}\right](1+z)^2e^{-\tau_{\rm es}},
\end{equation}
where $d\eta=(1+z)cdt$, $\tau_{\rm es}$ is the optical depth from
redshift 0 to $z$ and the $Q^{\rm m}$ are mass filling factors.  It is commonly 
assumed that the visibility function is
peaked during the recombination era, with a second, lower amplitude, but
broader peak coinciding with the epoch of overlap of HII regions. However
we have found that overlap may have occurred on multiple occasions, and that
many combinations of the parameters $z_{\rm tran}$ and $f_{\rm esc}f_\star$
predict significant reionization without overlap ({\it failed overlap}) at
high redshift due to the early presence of metal-free stars. During these
failed overlaps, examples of which are shown in figures~\ref{fig2} and
\ref{fig3}, the ionization fraction can be significant. Following a failed
overlap at $z_{\rm tran}$, recombinations
dominate the evolution and the ionization fraction drops before further
reionization ensues due to normal stars and quasars. The higher cosmic
density at early times might more than compensate for the smaller
ionization fraction, so that the peak of the visibility function does not
coincide with the epoch of reionization.

\begin{figure*}[htbp]
\epsscale{1}
\plotone{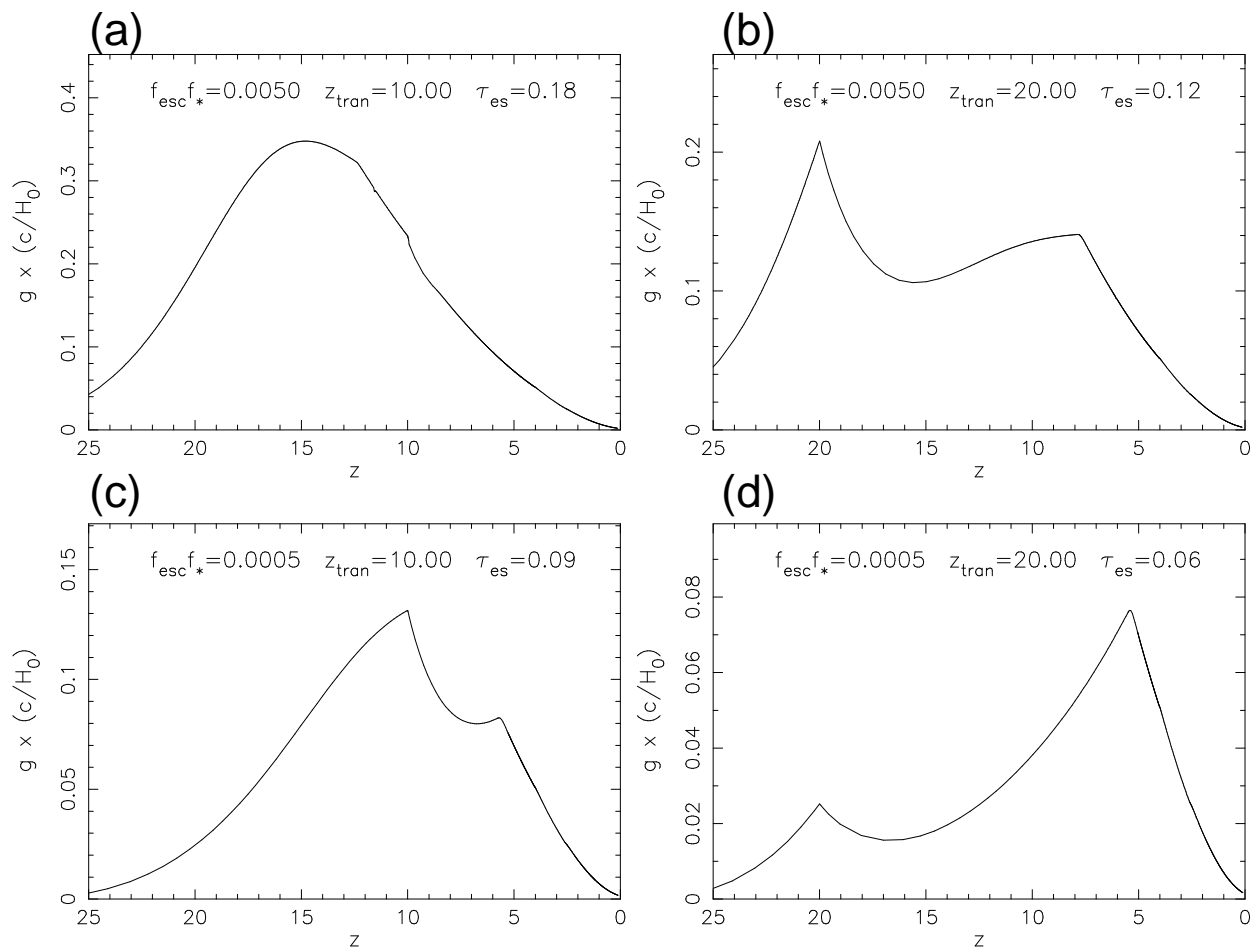}
\caption{\label{fig8} Sample visibility functions (in units of
$H_{0}/c$). The four cases correspond to the sample reionization histories 
(assuming case A star-formation) plotted in figure~\ref{fig2}. }
\end{figure*}

This is indeed what we find. Figures~\ref{fig7} and \ref{fig8} show visibility 
functions corresponding
to the sample reionization histories shown in figures~\ref{fig2} and
\ref{fig3}. Generally, if the early metal-free stars succeed in achieving
reionization, then the peak in the visibility function coincides with this
overlap epoch (e.g. case A in figures~\ref{fig2} and \ref{fig3}).  However
the optical depth $\tau_{\rm es}$ is large and many of these cases are already
excluded by CMB measurements. Of the 8 cases shown in figures~\ref{fig2} and \ref{fig3}, 
those that exhibit failed overlap have visibility peaks that coincide
with this failed overlap, rather than with the actual reionization epoch.
{\it Thus, the visibility function may probe the nature of the early
generation of stars rather than the reionization epoch itself.}

The reionized gas during a failed overlap will be in bubbles. Gruzinov \&
Hu~(1998) have pointed out that a universe which is ionized in patches
induces additional CMB anisotropies on arc-minute scales. Thus, patchy
reionization due to a failed overlap at high redshifts will modify the
shape of the small-scale power spectrum of CMB anisotropies.  However, the
amplitude of the added secondary anisotropies was predicted to be at a
level well below the feature in the spectrum detected more recently by the
Cosmic Background Imager (Padin et al~2001). This feature was explained by
Bond et al.~(2002) as being due to the foreground Sunyaev-Zeldovich effect
from X-ray clusters.

%FIGURE 8

%FIGURE 9
\begin{figure*}[htbp]
\epsscale{1}
\plotone{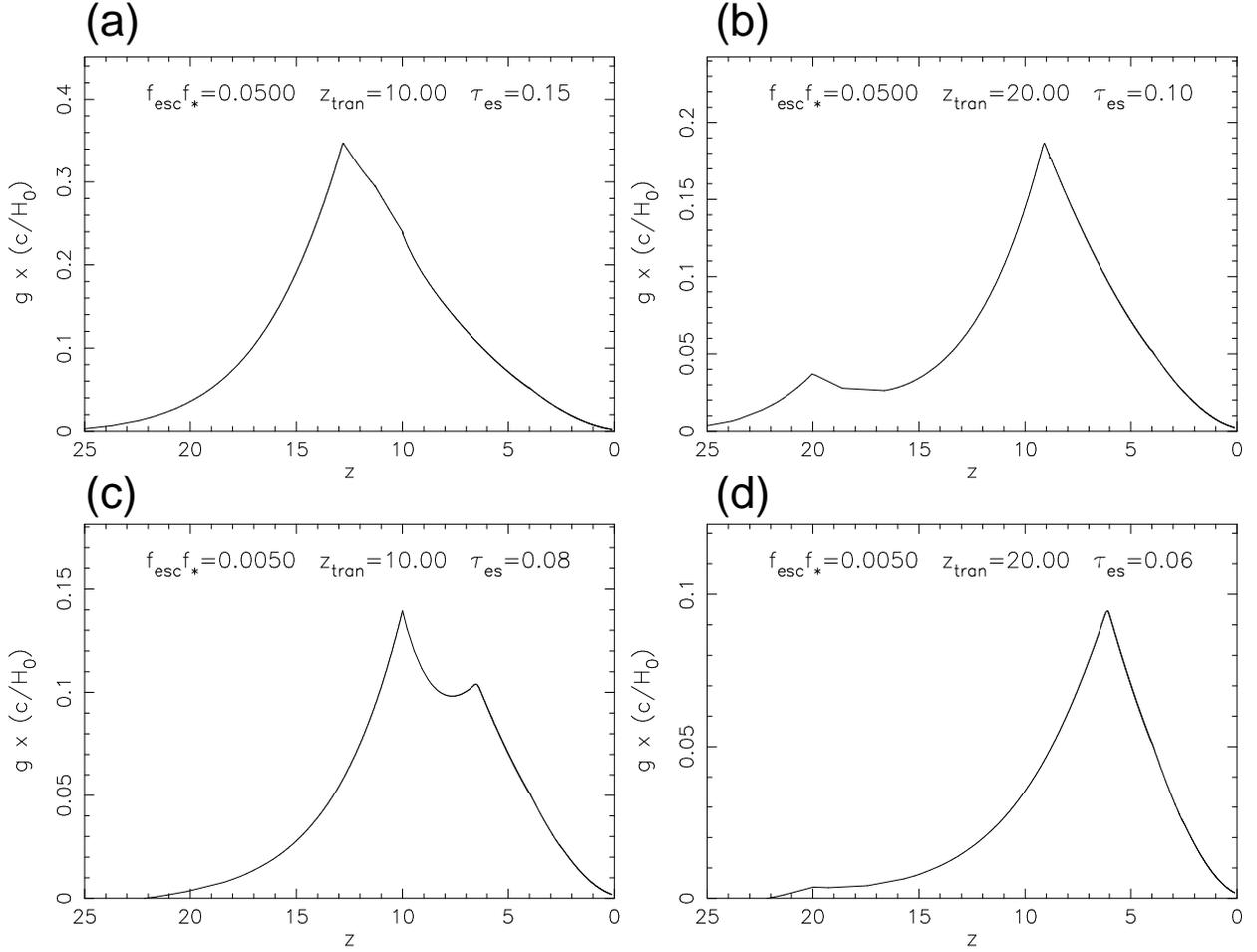}
\caption{\label{fig9} Sample visibility functions (in units of
$H_{0}/c$). The four cases correspond to the sample reionization histories 
(assuming case B star-formation) plotted in figure~\ref{fig3}. }
\end{figure*}

\section{Justification for Step-Function Approach to the 
Evolution in the Mode of Star Formation}
\label{justification}
As mentioned in \S~\ref{stars} we have approximated the evolution in
the mode of star formation as a step function. That is, we have
assumed that the production rate of ionizing photons in the universe
is transformed sharply at $z_{\rm tran}$ from being dominated by
massive, metal-free stars to being dominated by metal-rich stars with
a Scalo IMF.  In this section we justify this approach by comparing
its results to those obtained with a scheme involving gradual
metal enrichment of the IGM as described in \S~\ref{stars} by
equation~(\ref{gradualenrich}).

%FIGURE 10
\begin{figure*}[htbp]
\epsscale{1}
\plotone{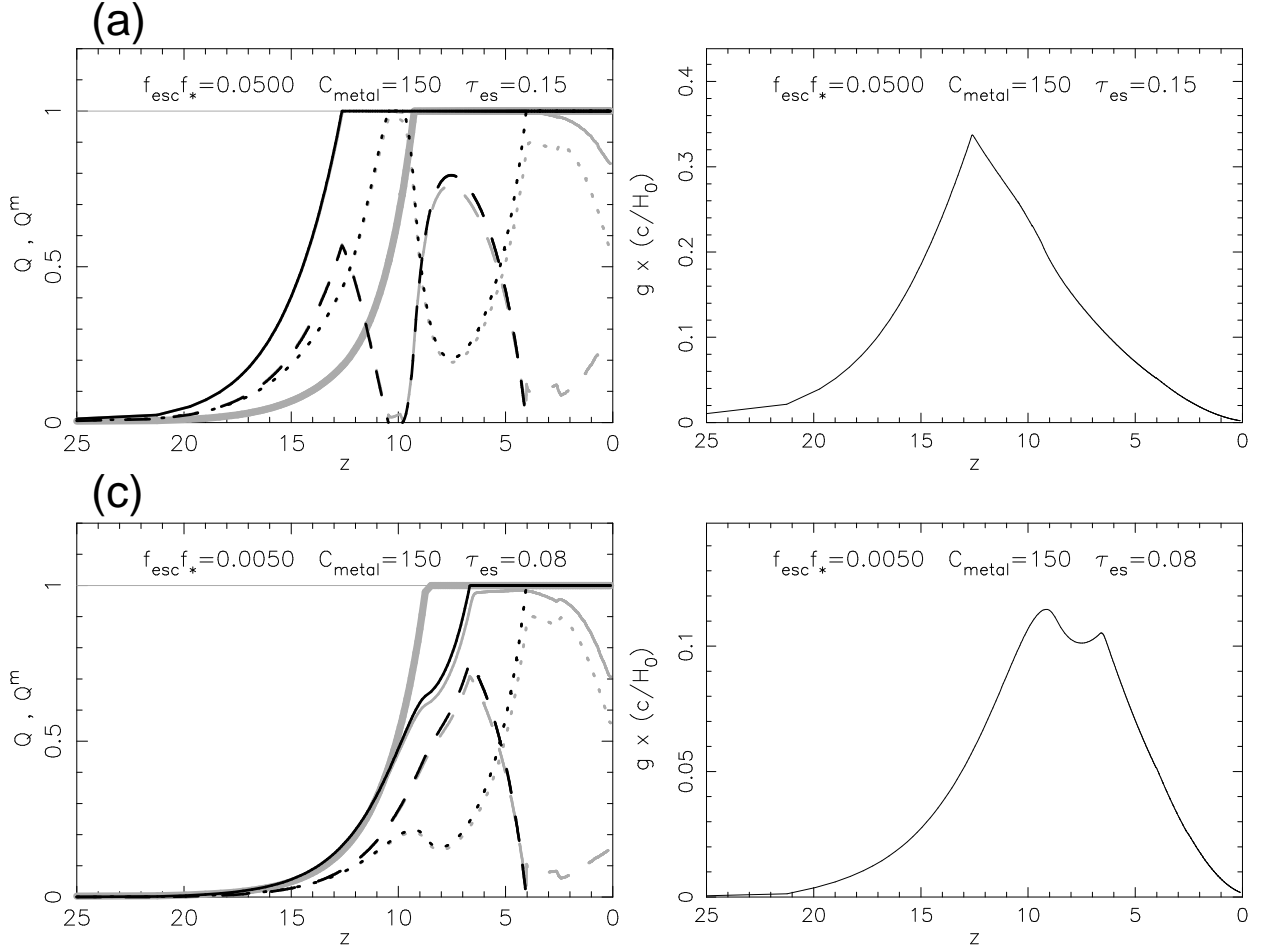}
\caption{\label{fig10} Results for reionization histories computed
assuming continuous enrichment of the IGM rather than sudden
enrichment at $z_{\rm tran}$. The results should be compared with
panels a) and c) of Figures~\ref{fig3} and \ref{fig9}.  Left: Sample
reionization histories assuming case B star formation. Dark lines
represent the evolution of the volume filling factors $Q_{\rm H^+}$,
$Q_{\rm He^+}$ and $Q_{\rm He^{++}}$, while the light lines show the
evolution of the fractions of ionized mass in the universe $Q^{\rm
m}_{\rm H^+}$, $Q^{\rm m}_{\rm He^+}$ and $Q^{\rm m}_{\rm
He^{++}}$. The solid, dashed and dotted lines refer to ${\rm H^+}$,
$\rm He^+$ and $\rm He^{++}$, respectively.  Right: The corresponding
visibility functions (in units of $H_{0}/c$). }
\end{figure*}

Figure~\ref{fig10} shows two sample reionization histories computed
using the gradual enrichment scheme, together with the resulting
visibility function. The examples correspond to panels a) and c) for
Case B star formation in Figure~\ref{fig3}. The thick grey line in
these panels shows the fraction of star formation resulting in a
Scalo~(1998) mass function as a function of redshift. \ In each of
these cases the value of $C_{\rm metal}$ (listed in the figure)
produces around 50\% Scalo~(1998) IMF star formation at the previously
used redshift $z_{\rm tran}$. Note the different values of $C_{\rm
metal}$, and the different histories of the collapsed mass-fraction.
The features discussed for the corresponding examples in
Figures~\ref{fig3} and \ref{fig9} are still present in
Figure~\ref{fig10}. These include the double overlap of He~III in a),
and the non-coincidence of the overlap redshift of HII and the peak in
the visibility function in b). Furthermore, we find unchanged values
of $\tau_{\rm es}$. These calculations appear to justify our use of a
single parameter $z_{\rm tran}$ to characterize the reionization
histories and visibility functions.

\section{Conclusion}

We have explored the reionization histories of hydrogen and helium due to
stars and quasars. The results were analyzed as a function of the two free
parameters in our model, namely: (i) the transition redshift, $z_{\rm
tran}$, above which the stellar population is dominated by massive, zero
metallicity stars; and (ii) the product of the escape fraction of ionizing
photons and the star-formation efficiency, $f_{\rm esc} f_\star$.  The
quasar model was not varied since it provides an excellent fit to all
existing data on the luminosity function of quasars up to redshift $z\sim
6$ (Wyithe \& Loeb 2002).

Figures~\ref{fig2} and \ref{fig3} show sample reionization histories for
different choices of $z_{\rm tran}$ and $f_{\rm esc}f_\star$. We find that
a wide range of $z_{\rm tran}\ga 6$ is allowed for $f_{\rm esc}f_\star\sim
2\times10^{-3}$ if the star-formation efficiency does not depend on galaxy mass
(figure~\ref{fig4}) or $f_{\rm esc}f_\star\sim 10^{-2}$ if the efficiency is
suppressed in low mass galaxies (figure~\ref{fig5}). This wide range
satisfies the constraints that overlap of the HII regions must be achieved
by $z\sim 6$ (Fan et al 2002) and that the optical depth for electron
scattering must be limited to $\tau_{\rm es}<0.18$.  The allowed range leads
generically to $\tau_{\rm es}\ga 7\%$ (see figures~\ref{fig4} and \ref{fig5}).
The MAP satellite is expected to have sufficient sensitivity to detect
values of $\tau_{\rm es}$ as small as $\sim 5\%$ (Kaplinghat et al. 2002),
well below this range of expected values.

A major fraction of the allowed range of $f_{\rm esc}f_{\star}$ and $z_{\rm
tran}$ leads to an early peak in the ionized fraction due to metal-free
stars at high redshifts.  Often this peak results in a small but
significant filling factor that is subsequently reduced (temporarily) due
to recombination. In a restricted range of the allowed parameter values, we
find that either hydrogen or helium experience two overlap epochs,
separated by recombination (see shaded regions in figures~\ref{fig6} and
\ref{fig8}). That helium might have been reionized twice due to the presence
of early very massive stars was previously pointed out by Oh et al.~(2001).
The first overlap phase is caused by the population of
zero-metallicity, massive stars and the second is dominated by the quasars
for helium or by stars and quasars for hydrogen.  Even if early overlap is
not achieved, the peak in the visibility function for scattering of the CMB
often coincides with the early partial ionization peak rather than with the
actual reionization epoch (see figures~\ref{fig7} and \ref{fig9}). The resulting
value of $\tau_{\rm es}$ is therefore larger than expected based the
reionization redshift alone.  Thus, the CMB visibility function may be
probing the nature of the early generation of stars rather than the
reionization epoch itself.

Future CMB experiments, such as MAP and Planck, will provide tighter
constraints on $\tau_{\rm es}$ and will reduce the range of allowed
$f_{\rm esc}f_\star$ in our model (see figures~\ref{fig4} and \ref{fig5}). Any
additional observational data on the composition of the stellar population
or the abundance of HeI or HeII at $z\ga 4$ will provide stronger lower
limits on $z_{\rm tran}$.

\acknowledgements 

We thank Rennan Barkana and an anonymous referee for useful detailed 
comments on the manuscript.
JSBW is supported by a Hubble Fellowship grant from the Space Telescope
Science Institute, which is operated by the Association of Universities for
Research in Astronomy, Inc., under NASA contract NAS 5-26555.  AL
acknowledges generous support from the Institute for Advanced Study at
Princeton and the John Simon Guggenheim Memorial Fellowship.  This work was
also supported in part by NSF grant AST-0071019.

\newpage
\begin{appendix}

\section{Equations Describing Reionization of Hydrogen and Helium in a 
Clumpy Universe}

The filling factors $Q_{\rm H^+}$, $Q_{\rm He^+}$ and $Q_{\rm He^{++}}$, and 
ionized mass fractions $F_{\rm H^+}$, $F_{\rm He^+}$ and $F_{\rm He^{++}}$ 
evolve according to
six coupled ODEs. It is convenient to write the ODEs and constraints in
terms of the following quantities:
\begin{eqnarray}
\nonumber S_{\rm CF}^{\rm H^+} &=& \frac{1}{n_{\rm H}^{\rm 0}}\frac{dn_{\rm
\gamma}^{\rm H^+}}{dz}\\ \nonumber S_{\rm CF}^{\rm He^{+}} &=&
\frac{1}{n_{\rm He}^{\rm 0}}\frac{dn_{\rm \gamma}^{\rm He^{+}}}{dz}\\
\nonumber S_{\rm CF}^{\rm He^{++}} &=& \frac{1}{n_{\rm He}^{\rm
0}}\frac{dn_{\rm \gamma}^{\rm He^{++}}}{dz}\\ \nonumber S_{\rm C}^{\rm H^+}
&=& \frac{1}{n_{\rm H}^{\rm 0}F_{\rm H^{+}}}\frac{dn_{\rm \gamma}^{\rm
H^+}}{dz}\\ \nonumber S_{\rm C}^{\rm He^+} &=& \frac{1}{n_{\rm He}^{\rm
0}F_{\rm He^+}}\frac{dn_{\rm \gamma}^{\rm He^+}}{dz}\\ \nonumber S_{\rm
C}^{\rm He^{++}} &=& \frac{1}{n_{\rm He}^{\rm 0}F_{\rm
He^{++}}}\frac{dn_{\rm \gamma}^{\rm He^{++}}}{dz}\\ \nonumber S_{\rm
KF}^{\rm H^+} &=& \alpha_{\rm B}^{\rm H^+}\frac{R^{\rm H^+}}{a^3}n_{\rm
e}^{\rm H^+}\frac{dt}{dz}\\ \nonumber S_{\rm KF}^{\rm He^+} &=& \alpha_{\rm
B}^{\rm He^+}\frac{R^{\rm He^+}}{a^3}n_{\rm e}^{\rm He^+}\frac{dt}{dz}\\
\nonumber S_{\rm KF}^{\rm He^{++}} &=& \alpha_{\rm B}^{\rm
He^{++}}\frac{R^{\rm He^{++}}}{a^3}n_{\rm e}^{\rm He^{++}}\frac{dt}{dz}\\
\nonumber S_{\rm K}^{\rm H^+} &=& \left(\alpha_{\rm B}^{\rm
H^+}\frac{R^{\rm H^+}}{a^3}n_{\rm e}^{\rm H^+}\frac{dt}{dz} + \frac{dF_{\rm
H^+}}{dz}\right)\frac{Q_{\rm H^+}}{F_{\rm H^+}}\\ \nonumber S_{\rm K}^{\rm
He^+} &=& \left(\alpha_{\rm B}^{\rm He^+}\frac{R^{\rm He^+}}{a^3}n_{\rm
e}^{\rm He^+}\frac{dt}{dz} + \frac{dF_{\rm He^+}}{dz}\right)\frac{Q_{\rm
He^+}}{F_{\rm He^+}}\\ S_{\rm K}^{\rm He^{++}} &=&
\left(\alpha_{\rm B}^{\rm He^{++}}\frac{R^{\rm He^{++}}}{a^3}n_{\rm e}^{\rm
He^{++}}\frac{dt}{dz} + \frac{dF_{\rm He^{++}}}{dz}\right)\frac{Q_{\rm
He^{++}}}{F_{\rm He^{++}}}.
\end{eqnarray}
In the above definitions, the $S_{\rm C}$ and $S_{\rm CF}$ are source terms
describing the ionizing radiation field, while the $S_{\rm K}$ and $S_{\rm
KF}$ are sink terms describing the recombination rate. In the recombination
terms, the $n_{\rm e}$ are the electron densities available for
recombination (equal to the mean cosmological density multiplied by the
volume filling factor and the mass fraction). These are therefore
\begin{eqnarray}
\nonumber n_{\rm e}^{\rm H^+} &=& \left[n_{\rm H}^{\rm 0} + n_{\rm He}^{\rm
0}\frac{Q_{\rm He^+}F_{\rm He^{+}}+2Q_{\rm He^{\rm ++}}F_{\rm
He^{++}}}{Q_{\rm H^+}F_{\rm H^{+}}}\right]\\ \nonumber n_{\rm e}^{\rm He^+}
&=& \left[n_{\rm H}^{\rm 0} + n_{\rm He}^{\rm 0}\right]\\ \nonumber \mbox{and}\\
n_{\rm e}^{\rm He^{++}} &=& \left[n_{\rm H}^{\rm 0} + 2n_{\rm He}^{\rm 0}\right].
\end{eqnarray}

We write the generalization of equations~(\ref{postoverlap}) and
(\ref{preoverlap}) for the three species in terms of the above
quantities. There are 3 cases corresponding to the period before HII
regions overlap, the period following the overlap of HII regions but
preceding the overlap of HeII regions, and the period following the
overlap of HeII regions:

\noindent$\bullet${\bf CASE I) Pre Hydrogen Overlap: If $Q_{\rm H^{+}}<1$
or $F_{\rm H^{+}}<F_{\rm M}(\Delta_{\rm crit})$}

Prior to the overlap of HII regions, the mass fractions $F$ and their
derivatives are set by the critical value of $\Delta_{\rm crit}$ chosen as the
ionization density threshold. The filling factors $Q_{\rm H^+}$, $Q_{\rm He^+}$
and $Q_{\rm He^{++}}$ evolve according to a
generalization of equation~(\ref{postoverlap}). Hence
\begin{eqnarray}
\label{ODE1}
\nonumber \frac{dF_{\rm H^{+}}}{dz} &=& \frac{d}{dz}F_{\rm M}(\Delta_{\rm
crit}) \hspace{20mm} F_{\rm H^{+}}=F_{\rm M}(\Delta_{\rm crit}) \\ \nonumber
\frac{dF_{\rm He^{+}}}{dz} &=& \frac{d}{dz}F_{\rm M}(\Delta_{\rm crit})
\hspace{20mm} F_{\rm He^{+}}=F_{\rm M}(\Delta_{\rm crit}) \\ \nonumber
\frac{dF_{\rm He^{++}}}{dz} &=& \frac{d}{dz}F_{\rm M}(\Delta_{\rm crit})
\hspace{20mm} F_{\rm He^{++}}=F_{\rm M}(\Delta_{\rm crit}) \\ \nonumber
\frac{dQ_{\rm H^+}}{dz}&=&S_{\rm C}^{\rm H^{+}} + \frac{1}{n_{\rm H}^{\rm
0}F_{\rm H^{+}}}\left[\Delta\frac{dn_{\rm \gamma}^{\rm
He^+}}{dz}+\Delta\frac{dn_{\rm \gamma}^{\rm He^{++}}}{dz}\right] - S_{\rm
K}^{\rm H^{+}}\\ \nonumber \frac{dQ_{\rm He^{\rm +}}}{dz}&=& S_{\rm C}^{\rm
He^{+}} - \frac{1}{n_{\rm He}^{\rm 0}F_{\rm He^{+}}}\Delta\frac{dn_{\rm
\gamma}^{\rm He^{+}}}{dz}- S_{\rm K}^{\rm He^{+}} -\frac{dQ_{\rm He^{\rm
++}}}{dz}\\ \frac{dQ_{\rm He^{\rm ++}}}{dz}&=&S_{\rm C}^{\rm He^{++}} -
\frac{1}{n_{\rm He}^{\rm 0}F_{\rm He^{++}}}\Delta\frac{dn_{\rm \gamma}^{\rm
He^{++}}}{dz}- S_{\rm K}^{\rm He^{++}}.
\end{eqnarray}
In the above, $\Delta\frac{dn_{\rm \gamma}^{\rm He^+}}{dz}$ and
$\Delta\frac{dn_{\rm \gamma}^{\rm H^{++}}}{dz}$ are the excess ionizing
photon rates having frequencies $5.94\times10^{15}<\nu<1.31\times10^{16}$Hz
for HeI ionization, and $\nu>1.31\times10^{16}$Hz for HeII
ionization. The values of these excess source terms are determined from the
constraints on the relative evolution of the different ionization
fronts. As these are excess rates, their values must be positive.
 
If $Q_{\rm He^{++}}<Q_{\rm H^{+}}$ then $Q_{\rm He^{+}}$ is limited to be
smaller than $Q_{\rm H^{+}}-Q_{\rm He^{++}}$ since the $\rm He^{+}$ front
cannot propagate beyond the H$^{+}$ front. Therefore, if $Q_{\rm He^{+}}=
Q_{\rm H^{+}}-Q_{\rm He^{++}}$ we require $\frac{dQ_{\rm
He^{+}}}{dz}=\frac{dQ_{\rm H^{+}}}{dz}-\frac{dQ_{\rm He^{++}}}{dz}$, and
since $\Delta\frac{dn_{\rm \gamma}^{\rm He^{++}}}{dz}=0$ in this case we
find
\begin{eqnarray}
\Delta\frac{dn_{\rm \gamma}^{\rm He^{+}}}{dz}&=&\left(\frac{1}{n_{\rm
He}^{\rm 0}F_{\rm He^{+}}}+\frac{1}{n_{\rm H}^{\rm 0}F_{\rm
H^{+}}}\right)^{-1} \left[(S_{\rm C}^{\rm He^{+}}-S_{\rm C}^{\rm H^{+}}) -
(S_{\rm K}^{\rm He^{+}}-S_{\rm K}^{\rm H^{+}}) \right].
\end{eqnarray}
In addition to the above constraint, the $\rm He^{++}$ filling factor $Q_{\rm
He^{++}}$ is limited to be less than $Q_{\rm H^{+}}$. Therefore, if $Q_{\rm
He^{++}}= Q_{\rm H^{+}}$ we require $\frac{dQ_{\rm
He^{++}}}{dz}=\frac{dQ_{\rm H^{+}}}{dz}$ so that the fronts propagate at
the same rate, and $\frac{dQ_{\rm He^{+}}}{dz}=0$ since the He$^+$ front
cannot propagate if the HeII region is filling all of the available
volume. These constraints yield
\begin{eqnarray}
\nonumber \Delta\frac{dn_{\rm \gamma}^{\rm He^{+}}}{dz}&=&\frac{n_{\rm
He}^{\rm 0}F_{\rm He^{+}}n_{\rm H}^{\rm 0}F_{\rm H^{+}}}{2n_{\rm He}^{\rm
0}F_{\rm He^{++}}+n_{\rm H}^{\rm 0}F_{\rm H^{+}}}\left[ (S_{\rm C}^{\rm
He^{++}}-S_{\rm C}^{\rm H^{+}}) - (S_{\rm K}^{\rm He^{++}}-S_{\rm K}^{\rm
H^{+}})\right]\\ \nonumber &-&n_{\rm He}^{\rm 0}F_{\rm He^{+}}\frac{n_{\rm
He}^{\rm 0}F_{\rm He^{++}}+n_{\rm H}^{\rm 0}F_{\rm H^{+}}}{2n_{\rm He}^{\rm
0}F_{\rm He^{++}}+n_{\rm H}^{\rm 0}F_{\rm H^{+}}}\left[ (S_{\rm C}^{\rm
He^{++}}-S_{\rm C}^{\rm He^{+}}) - (S_{\rm K}^{\rm He^{++}}-S_{\rm K}^{\rm
He^{+}})\right]\\ \nonumber &&\mbox{and}\\ \Delta\frac{dn_{\rm \gamma}^{\rm
He^{++}}}{dz}&=&n_{\rm He}^{\rm 0}F_{\rm He^{++}}\left[ (S_{\rm C}^{\rm
He^{++}}-S_{\rm C}^{\rm He^{+}}) - (S_{\rm K}^{\rm He^{++}}-S_{\rm K}^{\rm
He^{+}}) + \frac{1}{n_{\rm He}^{\rm 0}F_{\rm He^{+}}}\Delta\frac{dn_{\rm
\gamma}^{\rm He^{+}}}{dz}\right].
\end{eqnarray}

\noindent$\bullet${\bf CASE II) Post Hydrogen Overlap: If $Q_{\rm H^{+}}=1$
and $F_{\rm H^{+}}>F_{\rm M}(\Delta_{\rm crit})$ and ($Q_{\rm He^{++}}<1$ or
$F_{\rm He^{++}}<F_{\rm M}(\Delta_{\rm crit})$).}

Following the overlap of HII regions, the mass fraction of hydrogen $F_{\rm
H^+}$ is free to evolve according to equation~(\ref{preoverlap}) (with
modified source term to account for the excess ionizing photons). The
filling factors $Q_{\rm H^+}$, $Q_{\rm He^+}$ and $Q_{\rm He^{++}}$ 
again evolve according to a generalization of
equation~(\ref{postoverlap}). Note that if $Q_{\rm i}=1$, substitution of
equation~(\ref{postoverlap}) into equation~(\ref{preoverlap}) yields
$\frac{dQ_{\rm i}}{dz}=0$. Thus, post HII overlap the value of $Q_{\rm
H^+}$ remains at unity. However by continuing to follow all the equations we
allow for the possibility of a recombination epoch following an early
reionization. We have
\begin{eqnarray}
\label{ODE2}
\nonumber \frac{dF_{\rm H^{+}}}{dz} &=& S_{\rm CF}^{\rm H^{+}} +
\frac{1}{n_{\rm H}^{\rm 0}}\Delta\frac{dn_{\rm \gamma}^{\rm He^+}}{dz}-
S_{\rm KF}^{\rm H^{+}}\\ \nonumber \frac{dF_{\rm He^{+}}}{dz} &=&
\frac{d}{dz}F_{\rm M}(\Delta_{\rm crit}) \hspace{20mm} F_{\rm He^{+}}=F_{\rm
M}(\Delta_{\rm crit}) \\ \nonumber \frac{dF_{\rm He^{++}}}{dz} &=&
\frac{d}{dz}F_{\rm M}(\Delta_{\rm crit}) \hspace{20mm} F_{\rm He^{++}}=F_{\rm
M}(\Delta_{\rm crit}) \\ \nonumber \frac{dQ_{\rm H^+}}{dz}&=&S_{\rm C}^{\rm
H^{+}} + \frac{1}{n_{\rm H}^{\rm 0}F_{\rm H^{+}}}\Delta\frac{dn_{\rm
\gamma}^{\rm He^+}}{dz} - S_{\rm K}^{\rm H^{+}}\\ \nonumber \frac{dQ_{\rm
He^{\rm +}}}{dz}&=& S_{\rm C}^{\rm He^{+}} - \frac{1}{n_{\rm He}^{\rm
0}F_{\rm He^{+}}}\Delta\frac{dn_{\rm \gamma}^{\rm He^{+}}}{dz}- S_{\rm
K}^{\rm He^{+}} -\frac{dQ_{\rm He^{\rm ++}}}{dz}\\ \frac{dQ_{\rm He^{\rm
++}}}{dz}&=&S_{\rm C}^{\rm He^{++}}- S_{\rm K}^{\rm He^{++}}.
\end{eqnarray}
As before, the values of the excess ionizing photon rates are determined
from constraints on the co-evolution of the ionization fronts. In the case
where $Q_{\rm He^{++}}<Q_{\rm H^{+}}$, $Q_{\rm He^{+}}$ is limited to be
smaller than $Q_{\rm He^{+}}=Q_{\rm H^{+}}-Q_{\rm He^{++}}$. However as
mentioned above, if $Q_{\rm H^{+}}=1$ then the above equations imply
$\frac{dQ_{\rm H^{+}}}{dz}=0$. We therefore have $\frac{dQ_{\rm
He^{+}}}{dz}=-\frac{dQ_{\rm He^{++}}}{dz}$, and since $\Delta\frac{dn_{\rm
\gamma}^{\rm He^{++}}}{dz}=0$ in this case we find
\begin{equation}
\Delta\frac{dn_{\rm \gamma}^{\rm He^{+}}}{dz}=n_{\rm He}^{\rm 0}F_{\rm
He^{+}}\left[ S_{\rm C}^{\rm He^{+}}-S_{\rm K}^{\rm He^{+}}\right].
\end{equation}

\noindent$\bullet${\bf CASE III) Post Helium Overlap: If $Q_{\rm
He^{++}}=1$ and $F_{\rm He^{++}}\ge F_{\rm M}(\Delta_{\rm crit})$.}

Finally, following the overlap of HeII regions, the mass fractions $F_{\rm
H^+}$, $F_{\rm He^{+}}$ and $F_{\rm He^{++}}$ are free to evolve according to
equation~(\ref{preoverlap}) (with modified source terms to account for the
excess ionizing photons). The filling factors $Q_{\rm H^+}$, $Q_{\rm He^+}$ and
$Q_{\rm He^{++}}$ still evolve according to
generalizations of equation~(\ref{postoverlap}). At the point of He$^{++}$
overlap our formalism demands that there is no He$^+$ in the universe 
since $Q_{\rm He^+}=0$. However, following He$^{++}$ overlap the He$^{+}$ front
(which must lead the He$^{++}$ front) is free to propagate into the denser IGM.
Therefore, following the overlap of He$^{++}$, $F_{\rm He^+}$ becomes the fraction 
of helium in the universe (equal to zero initially) that is singly ionized. 
The recombination rate $R^{\rm He^+}$ is calculated 
\begin{equation}
R^{\rm He^+}=\int_{\Delta_{\rm i}^{\rm He^{++}}}^{\Delta_{\rm i}^{\rm He^{+}}}d\Delta\frac{dP}{d\Delta}\Delta^2,
\end{equation}
where $F_{\rm He^{++}} = \int_{0}^{\Delta_{\rm i}^{\rm He^{++}}}d\Delta\frac{dP}{d\Delta}\Delta$ and $F_{\rm He^{++}}+F_{\rm He^{+}} = \int_{0}^{\Delta_{\rm i}^{\rm He^{+}}}d\Delta\frac{dP}{d\Delta}\Delta$. As before, if $Q_{\rm
i}=1$ substitution of equation~(\ref{postoverlap}) into
equation~(\ref{preoverlap}) yields $\frac{dQ_{\rm i}}{dz}=0$. The filling
factors $Q_{\rm H^+}$ and $Q_{\rm He^{++}}$ therefore remain at unity. We have
\begin{eqnarray}
\label{ODE3}
\nonumber\frac{dF_{\rm H^{+}}}{dz} &=& S_{\rm CF}^{\rm H^{+}} + \frac{1}{n_{\rm
H}^{\rm 0}}\left[\Delta\frac{dn_{\rm \gamma}^{\rm He^+}}{dz} +
\Delta\frac{dn_{\rm \gamma}^{\rm He^{++}}}{dz}\right]- S_{\rm KF}^{\rm
H^{+}}\\ \nonumber \frac{dF_{\rm He^{+}}}{dz} &=& S_{\rm CF}^{\rm He^{+}} -
\frac{1}{n_{\rm He}^{\rm 0}}\Delta\frac{dn_{\rm \gamma}^{\rm He^+}}{dz} -
S_{\rm KF}^{\rm He^{+}} - \frac{dF_{\rm He^{++}}}{dz}\\ \nonumber
\frac{dF_{\rm He^{++}}}{dz} &=& S_{\rm CF}^{\rm He^{++}} - \frac{1}{n_{\rm
He}^{\rm 0}}\Delta\frac{dn_{\rm \gamma}^{\rm He^{++}}}{dz}- S_{\rm KF}^{\rm
He^{++}}\\ \nonumber \frac{dQ_{\rm H^+}}{dz}&=&S_{\rm C}^{\rm H^{+}} +
\frac{1}{n_{\rm H}^{\rm 0}F_{\rm H^{+}}}\left[\Delta\frac{dn_{\rm
\gamma}^{\rm He^+}}{dz}+\Delta\frac{dn_{\rm \gamma}^{\rm
He^{++}}}{dz}\right] - S_{\rm K}^{\rm H^{+}}\\ \nonumber \frac{dQ_{\rm
He^{\rm +}}}{dz}&=& S_{\rm C}^{\rm He^{+}} - \frac{1}{n_{\rm He}^{\rm
0}F_{\rm He^{+}}}\Delta\frac{dn_{\rm \gamma}^{\rm He^{+}}}{dz}- S_{\rm
K}^{\rm He^{+}} -\frac{dQ_{\rm He^{\rm ++}}}{dz}\\ \frac{dQ_{\rm He^{\rm
++}}}{dz}&=&S_{\rm C}^{\rm He^{++}}- \frac{1}{n_{\rm He}^{\rm 0}F_{\rm
He^{++}}}\Delta\frac{dn_{\rm \gamma}^{\rm He^{++}}}{dz}- S_{\rm K}^{\rm
He^{++}}.
\end{eqnarray}
The excess ionizing photon rates are determined as follows.  If $F_{\rm
He^{++}}<F_{\rm H^{+}}$ then $F_{\rm He^{+}}$ is limited to be smaller than
$F_{\rm He^{+}}=F_{\rm H^{+}}-F_{\rm He^{++}}$ since the He$^{+}$ front
cannot propagate beyond the H$^{+}$ front. Therefore, if $F_{\rm He^{+}}\ge
F_{\rm H^{+}}-F_{\rm He^{++}}$ we require $\frac{dF_{\rm
He^{+}}}{dz}=\frac{dF_{\rm H^{+}}}{dz}-\frac{dF_{\rm He^{++}}}{dz}$, and
since $\Delta\frac{dn_{\rm \gamma}^{\rm He^{++}}}{dz}=0$ in this case, we
find
\begin{equation}
\Delta\frac{dn_{\rm \gamma}^{\rm He^{+}}}{dz}=\left(\frac{1}{n_{\rm
He}^{\rm 0}}+\frac{1}{n_{\rm H}^{\rm 0}}\right)^{-1}\left[(S_{\rm CF}^{\rm
He^{+}}-S_{\rm CF}^{\rm H^{+}}) - (S_{\rm KF}^{\rm He^{+}}-S_{\rm KF}^{\rm
H^{+}}) \right].
\end{equation}
The $\rm He^{++}$ mass fraction $F_{\rm He^{++}}$ is limited to be less
than $F_{\rm H^{+}}$. Therefore, if $F_{\rm He^{++}}= F_{\rm H^{+}}$ we
require $\frac{dF_{\rm He^{++}}}{dz}=\frac{dF_{\rm H^{+}}}{dz}$ so that the
fronts propagate at the same rate, and $\frac{dF_{\rm He^{+}}}{dz}=0$ since
the He$^+$ front cannot grow if the $\rm He^{++}$ is filling all of the
available volume. These constraints yield
\begin{eqnarray}
\nonumber \Delta\frac{dn_{\rm \gamma}^{\rm He^{+}}}{dz}&=&\frac{n_{\rm
He}^{\rm 0}n_{\rm H}^{\rm 0}}{2n_{\rm He}^{\rm 0}+n_{\rm H}^{\rm 0}}\left[
(S_{\rm CF}^{\rm He^{++}}-S_{\rm CF}^{\rm H^{+}}) - (S_{\rm KF}^{\rm
He^{++}}-S_{\rm KF}^{\rm H^{+}})\right]\\ \nonumber &-&n_{\rm He}^{\rm
0}\frac{n_{\rm He}^{\rm 0}+n_{\rm H}^{\rm 0}}{2n_{\rm He}^{\rm 0}+n_{\rm
H}^{\rm 0}}\left[ (S_{\rm CF}^{\rm He^{++}}-S_{\rm CF}^{\rm He^{+}}) -
(S_{\rm KF}^{\rm He^{++}}-S_{\rm KF}^{\rm He^{+}})\right]\\\nonumber 
&&\mbox{and}\\ 
\Delta\frac{dn_{\rm \gamma}^{\rm He^{++}}}{dz}&=&n_{\rm He}^{\rm 0}\left[
(S_{\rm CF}^{\rm He^{++}}-S_{\rm CF}^{\rm He^{+}}) - (S_{\rm KF}^{\rm
He^{++}}-S_{\rm KF}^{\rm He^{+}}) + \frac{1}{n_{\rm He}^{\rm
0}}\Delta\frac{dn_{\rm \gamma}^{\rm He^{+}}}{dz}\right].
\end{eqnarray} 

\end{appendix}

\end{document}